\documentclass[sigconf]{acmart}

\usepackage{booktabs,caption}
\usepackage[flushleft]{threeparttable}
\usepackage{multirow}
\usepackage{enumitem}
\usepackage{bm}
\usepackage[normalem]{ulem}

\usepackage{marvosym}
\usepackage[ruled,linesnumbered]{algorithm2e}
\usepackage{amsmath}
\usepackage{braket}

\usepackage{subcaption}
\usepackage{makecell}
\usepackage{graphicx}
\usepackage{soul}
\usepackage{diagbox}
\usepackage{tabularx}
\usepackage{hyperref}
\hypersetup{
    colorlinks=true,
    linkcolor=black,
    filecolor=black,
    urlcolor=black,
    citecolor=black,
}

\usepackage{xspace}
\usepackage{adjustbox}

\AtBeginDocument{%
  \providecommand\BibTeX{{%
    \normalfont B\kern-0.5em{\scshape i\kern-0.25em b}\kern-0.8em\TeX}}}

\setcopyright{acmcopyright}
\copyrightyear{2022}
\acmYear{2022}

\acmConference[WWW '22] {Proceedings of the ACM Web Conference 2022}{April 25--29, 2022}{Virtual Event, Lyon, France.}
\acmBooktitle{Proceedings of the ACM Web Conference 2022 (WWW '22), April 25--29, 2022, Virtual Event, Lyon, France}
\acmPrice{15.00}
\acmISBN{978-1-4503-9096-5/22/04}
\acmDOI{10.1145/3485447.3512104}
\begin{document}

% \title{\emoji{telescope} Better to Look Wider: \\ Overlooked Contrastive Collaborative Filtering}
% \title[\emoji{telescope} Better to Look Wider: Overlooked Contrastive Collaborative Filtering]{\texorpdfstring{\emoji{telescope} Better to Look Wider: \\ Overlooked Contrastive Collaborative Filtering}{\emoji{telescope} Better to Look Wider: Overlooked Contrastive Collaborative Filtering}}
\title{Improving Graph Collaborative Filtering with Neighborhood-enriched Contrastive Learning}

\author{Zihan Lin$^{1\dagger}$, Changxin Tian$^{1\dagger}$, Yupeng Hou$^{2\dagger}$, Wayne Xin Zhao$^{2,3}$\textsuperscript{\Letter}}
\thanks{$\dagger$ Equal contribution. }
\thanks{\textsuperscript{\Letter} Corresponding author.}

\affiliation{
  \institution{$^1$School of Information, Renmin University of China, China}
  \institution{$^2$Gaoling School of Artificial Intelligence, Renmin University of China, China}
  \institution{$^3$Beijing Key Laboratory of Big Data Management and Analysis Methods, China}
  \country{}
}

\affiliation{
  \institution{ \{zhlin, tianchangxin, houyupeng\}@ruc.edu.cn, batmanfly@gmail.com}
  \country{ }
}
\newcommand{\ie}{\emph{i.e.,}\xspace}
\newcommand{\eg}{\emph{e.g.,}\xspace}
\newcommand{\aka}{\emph{a.k.a.,}\xspace}
\newcommand{\etal}{\emph{et al.}\xspace}
\newcommand{\paratitle}[1]{\vspace{1.5ex}\noindent\textbf{#1}}
\newcommand{\wrt}{w.r.t.\xspace}

\newcommand{\tian}[1]{\textcolor{black}{#1}}
\newcommand{\question}[1]{\textcolor{black}{#1}}
\newcommand{\ypcomment}[1]{\textcolor{black}{#1}}
\newcommand{\xin}[1]{\textcolor{blue}{#1}}

\newcommand{\modified}[1]{\textcolor{blue}{#1}}

\newcommand{\ignore}[1]{}

\newcommand{\our}{NCL~}

\renewcommand{\authors}{Zihan Lin, Changxin Tian, Yupeng Hou, Wayne Xin Zhao}
\renewcommand{\shortauthors}{Lin, Tian and Hou, et al.}

\begin{abstract}
  Recently, graph collaborative filtering methods have been proposed as an effective recommendation approach, which can  capture users' preference over items by modeling the user-item interaction graphs. 
  Despite the effectiveness, these methods suffer from data sparsity in real scenarios.
  In order to reduce the influence of data  sparsity, 
  contrastive learning is adopted in graph collaborative filtering for enhancing the performance. However, these methods typically construct the contrastive pairs by random sampling, which neglect the neighboring relations among users~(or items) and fail to fully exploit the potential of contrastive learning for recommendation.
  
  To tackle the above issue, we propose a novel contrastive learning approach, named \textbf{N}eighborhood-enriched \textbf{C}ontrastive \textbf{L}earning, named \textbf{NCL}, which explicitly incorporates the potential neighbors into contrastive pairs. 
  Specifically, we introduce the neighbors of a user~(or an item) from graph structure and semantic space respectively.
  For the structural neighbors on the interaction graph, we develop a novel structure-contrastive objective that regards users~(or items) and their structural neighbors as positive contrastive pairs. In implementation, the representations of users~(or items) and neighbors correspond to the outputs of different GNN layers.
  Furthermore, to excavate the potential neighbor relation in semantic space, we assume that users with similar representations are within the semantic neighborhood, and incorporate these semantic neighbors into the prototype-contrastive objective.
  The proposed NCL can be optimized with EM algorithm and generalized to apply to graph collaborative filtering methods. Extensive experiments on five public datasets demonstrate the effectiveness of the proposed NCL, notably with 26\% and 17\% performance gain over a competitive  graph collaborative filtering base model on the Yelp and Amazon-book datasets, respectively. 
  Our implementation code is available at: \url{https://github.com/RUCAIBox/NCL}.

\end{abstract}

\begin{CCSXML}
<ccs2012>
<concept>
<concept_id>10002951.10003317.10003347.10003350</concept_id>
<concept_desc>Information systems~Recommender systems</concept_desc>
<concept_significance>500</concept_significance>
</concept>
<concept>
<concept_id>10002951.10003227.10003351.10003269</concept_id>
<concept_desc>Information systems~Collaborative filtering</concept_desc>
<concept_significance>500</concept_significance>
</concept>
</ccs2012>
\end{CCSXML}

\ccsdesc[500]{Information systems~Recommender systems}
% \ccsdesc[500]{Information systems~Collaborative filtering}

\keywords{Recommender System, Collaborative Filtering, Contrastive Learning, Graph Neural Network}

\maketitle

\section{Introduction}
In the age of information explosion, recommender systems occupy an important position to discover users' preferences and deliver online services efficiently~\cite{ricci2011introduction}.
As a classic approach, collaborative filtering (CF)~\cite{sarwar2001item,he2017neural} is a fundamental technique that can produce effective recommendations from implicit feedback~(expression, click, transaction \etal). 
Recently, CF is further enhanced by the powerful  graph neural networks~(GNN)~\cite{he2020lightgcn,wang2019neural}, which models the interaction data as graphs (\eg  the user-item interaction graph) and then applies  GNN to learn effective node representations~\cite{he2020lightgcn,wang2019neural} for recommendation, called \emph{graph collaborative filtering}.

Despite the remarkable success, existing neural graph collaborative filtering methods still suffer from two major issues. 
Firstly, user-item interaction data is usually sparse or noisy, and it may not be able to learn reliable representations since the graph-based methods are potentially more vulnerable to data sparsity~\cite{wu2021self}. 
Secondly, existing GNN based CF approaches rely on explicit interaction links for learning node representations, while high-order relations or constraints (\eg user or item similarity) cannot be explicitly utilized for enriching the graph information, which has been shown essentially useful in recommendation tasks~\cite{sarwar2001item,wu2019neural,sun2019multi}. 
Although several recent studies leverage constative learning to alleviate the sparsity of interaction data~\cite{wu2021self,yao2020self}, they construct the 
contrastive pairs by randomly sampling nodes or corrupting subgraphs. It lacks consideration on how to construct more meaningful contrastive learning tasks 
tailored for the recommendation task~\cite{sarwar2001item,wu2019neural,sun2019multi}.

\begin{figure}[!ht]
    \centering
    \includegraphics[width=0.47\textwidth]{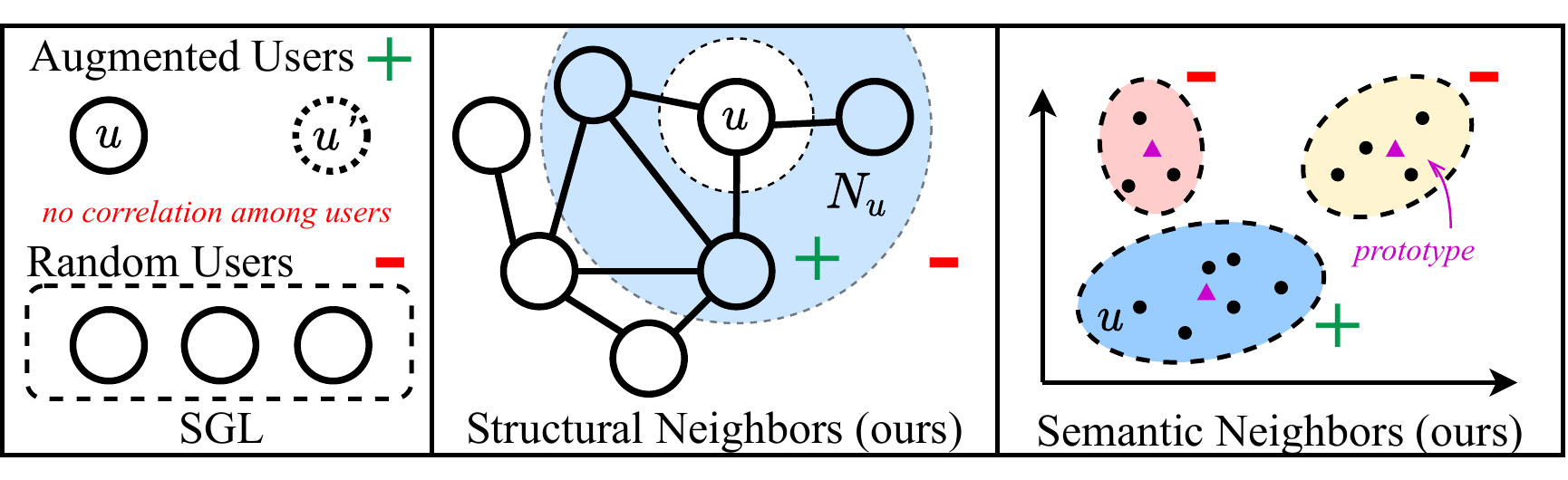}
    \vspace{-0.1in}
    \caption{Comparison of existing self-supervised learning approaches (\eg\ SGL~\cite{wu2021self}) that neglect the correlation among users (or items) and the proposed neighborhood-enriched contrastive learning approach (our approach).}
    \vspace{-0.15in}
    \label{fig:intro}
\end{figure}

Besides direct user-item interactions, there exist multiple kinds of potential relations (\eg user similarity) that are useful to the recommendation task, and 
we aim to design more effective constative learning approaches for leveraging such useful relations in neural graph collaborative filtering. 
Specially, we consider node-level relations \emph{w.r.t.} a user (or an item), which is more efficient than the graph-level relations. 
We characterize these additional relations as \emph{enriched neighborhood} of nodes, which can be defined in two aspects:  
(1) \textbf{structural neighbors} refer to  structurally connected nodes by high-order paths, and (2) \textbf{semantic neighbors} refer to semantically similar neighbors which may not be directly reachable on graphs. We aim to leverage these enriched node relations for improving the learning of node representations (\ie encoding user preference or item characteristics).

To integrate and model the enriched neighborhood, we propose \textbf{N}eighborhood-enriched \textbf{C}ontrastive \textbf{L}earning~(\textbf{NCL} for short), a model-agnostic constative learning framework for the recommendation. As introduced before, NCL constructs node-level contrastive objectives based on two kinds of extended neighbors. 
We present a comparison between NCL and existing constative learning methods in Figure~\ref{fig:intro}.
However, node-level contrastive objectives usually require pairwise learning for each node pair, which is time-consuming for large-sized neighborhoods. 
Considering the efficiency issue, we  learn a single representative embedding for each kind of neighbor, such that the constative learning for a node can be accomplished with two representative embeddings (either structural or semantic).

To be specific, for structural neighbors, we note that the outputs of $k$-th layer of GNN involve the aggregated information of $k$-hop neighbors. Therefore, we utilize the $k$-th layer output from GNN as the representations of $k$-hop neighbors for a node. We design a structure-aware contrastive learning objective that pulls the representations of a node (a user or item) and the representative embedding for its  structural neighbors.
For the semantic neighbors, we design a prototypical contrastive learning objective to capture the correlations between a node (a user or item) and its  prototype. Roughly speaking, a \emph{prototype} can be regarded as the centroid of the cluster of semantically similar neighbors in representation space.
Since the prototype is latent, we further propose to use an expectation-maximization~(EM) algorithm~\cite{moon1996expectation} to infer the prototypes. 
By incorporating these additional relations, our experiments show that it can largely improve the original GNN based approaches  (also better than existing constative learning methods) for implicit feedback recommendation.  
Our contributions can be summarized threefold:

$\bullet$ We propose a model-agnostic contrastive learning framework named NCL, which incorporates both structural and semantic neighbors for improving the neural graph collaborative filtering.

$\bullet$ We propose to learn representative embeddings for both kinds of neighbors, such that  the constative learning can be only performed between a node and the corresponding representative embeddings, which largely improves the algorithm efficiency.

$\bullet$ Extensive experiments are conducted on five public datasets, demonstrating that our approach is consistently better than a number of competitive baselines, including GNN and contrastive learning-based recommendation methods.

\section{Preliminary} \label{sec:Preliminary}
As the fundamental recommender system, collaborative filtering (CF) aims to recommend relevant items that users might be interested in  based on the observed implicit feedback (\eg~expression, click and transaction). 
Specifically, given the user set $\mathcal U=\{u\}$ and item set $\mathcal I=\{i\}$, the observed implicit feedback matrix is denoted as $ \textbf{R} \in \{0,1\}^{|\mathcal U| \times |\mathcal I|}$, where each entry $R_{u,i}=1$ if there exists an interaction between the user $u$ and item $i$, otherwise $R_{u,i}=0$.
Based on the interaction data $\textbf{R}$, the learned recommender systems can predict potential interactions for recommendation.
Furthermore, Graph Neural Network~(GNN) based collaborative filtering methods organize the interaction data $\textbf{R}$ as an interaction graph $\mathcal{G} = \{ \mathcal{V}, \mathcal{E} \}$, where $\mathcal{V} = \{ \mathcal{U} \cup \mathcal{I}\}$ denotes the set of nodes and $\mathcal{E}  =  \{(u,i) \;|\; u \in \mathcal{U} , i\in \mathcal{I}, R_{u,i} = 1 \} $ denotes the set of edges. 

In general, GNN-based collaborative filtering methods~\cite{wang2019neural,he2020lightgcn,wang2020disentangled} produce informative representations for users and items based on the aggregation scheme, which can be formulated to two stages: 
\begin{equation} 
\begin{aligned}
\bm z_u^{(l)} &= f_{\text{propagate}}(\{\bm z_v^{(l-1)}| v \in \mathcal{N}_{u} \cup \{v\} \}), \\
\bm z_u &= f_{\text{readout}}([\bm z_u^{(0)}, \bm z_u^{(1)},..., \bm z_u^{(L)}]),
\end{aligned}
\end{equation}
where $\mathcal{N}_{u}$ denotes the neighbor set of user $u$ in the interaction graph $\mathcal{G}$ and $L$ denotes the number of GNN layers. 
Here, $z_u^{(0)}$ is initialized by the learnable embedding vector $\mathbf e_u$. 
For the user $u$, the propagation function $f_{\text{propagate}}(\cdot)$ aggregates the $(l-1)$-th layer's representations of its neighbors to generate the $l$-th layer's representation $\mathbf z_u^{(l)}$.
After $l$ times iteratively propagation, the information of $l$-hop neighbors is encoded in $\mathbf z_u^{(l)}$. And the readout function $f_{\text{readout}}(\cdot)$ further summarizes all of the representations $[\bm z_u^{(0)}, \bm z_u^{(1)},..., \bm z_u^{(L)}]$ to obtain the final representations of user $u$  for recommendation. The informative representations of items can be obtained analogously.

\section{Methodology}
\begin{figure}[t]
    \centering
    \includegraphics[width=0.42\textwidth]{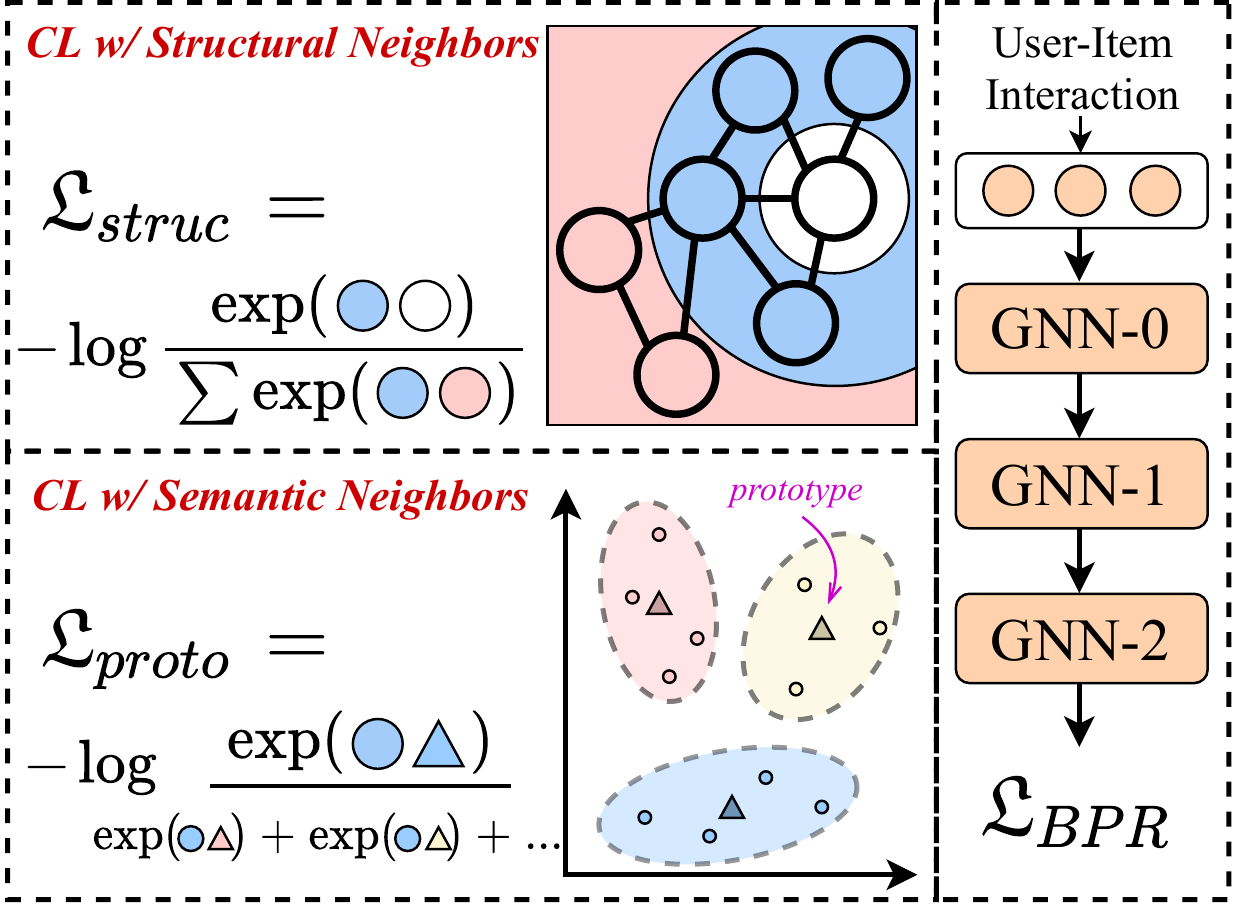}
    \caption{Overall framework of our proposed neighborhood-enriched contrastive collaborative filtering method.}
    \label{fig:overall}
\end{figure}

In this section, we introduce the proposed Neighborhood-enriched Contrastive Learning method in three parts. 
We first introduce the base graph collaborative filtering approach in Section~\ref{sec:GNN}, which outputs the final representations for recommendation along with the integrant representations for structural neighbors.
Then, we introduce the structure-contrastive strategies and prototype-contrastive strategies in Section~\ref{sec:Hop-Contrastive} and Section~\ref{sec:Prototype-Contrastive} respectively, which integrate the relation of neighbors into contrastive learning to coordinate with collaborative filtering properly.
Finally, we propose a multi-task learning strategy in Section~\ref{sec:Multi-task} and further present the theoretical analysis and discussion in Section~\ref{sec:Discussion}. The overall framework of \our is depicted in Figure~\ref{fig:overall}.

\subsection{Graph Collaborative Filtering BackBone} \label{sec:GNN}
{As mentioned in Section~\ref{sec:Preliminary}, GNN-based methods produce user and item representations by applying the propagation and prediction function on the interaction graph $\mathcal{G}$.
In NCL, we utilize GNN to model the observed interactions between users and items. 
Specifically, following LightGCN~\cite{he2020lightgcn}, we discard the nonlinear activation and feature transformation in the propagation function as:
\begin{equation} 
\begin{aligned}
\bm{z}_{u}^{(l+1)} &=\sum_{i \in N_{u}} \frac{1}{\sqrt{\left|\mathcal{N}_{u}\right|\left|\mathcal{N}_{i}\right|}} \bm{z}_{i}^{(l)}, \\
\bm{z}_{i}^{(l+1)} &=\sum_{u \in \mathcal{N}_{i}} \frac{1}{\sqrt{\left|\mathcal{N}_{i}\right|\left|\mathcal{N}_{u}\right|}} \bm{z}^{(l)} ,
\end{aligned}
\label{eq:graph_prop}
\end{equation}
After propagating with $L$ layers, we adopt the weighted sum function as the readout function to combine the representations of all layers and obtain the final representations as follows:
\begin{equation}
{\bm{z}}_{u}= \frac{1}{L+1} \sum_{l=0}^{L} {\bm{z}}_{u}^{(k)} , \;\;\;\; \;\;\;\;
{\bm{z}}_{i}= \frac{1}{L+1} \sum_{l=0}^{L} {\bm{z}}_{i}^{(k)},
\label{eq:layer_sum}
\end{equation}
where ${\bm{z}}_{u}$ and ${\bm{z}}_{i}$ denote the final representations of user $u$ and item $i$. 
With the final representations, we adopt inner product to predict how likely a user $u$ would interact with items $i$:
\begin{equation}
    \hat y_{u,i} =  \mathbf {z}_u^\top \mathbf {z}_i ,
    \label{eq:predict}
\end{equation}
where $\hat y_{u,i}$ is the prediction score of user $u$ and items $i$.
}

{To capture the information from interactions directly, we adopt Bayesian Personalized Ranking~(BPR) loss~\cite{rendle2009bpr}, which is a well-designed ranking objective function for recommendation. Specifically, BPR loss enforces the prediction score of the observed interactions higher than sampled unobserved ones. Formally, the objective function of BPR loss is as follows:
\begin{equation}
    \mathcal L_{BPR} = \sum_{(u,i,j)\in \mathcal{O}} -\log \sigma(\hat y_{u,i}-\hat y_{u,j}).
\end{equation}
where $\sigma(\cdot)$ is the sigmoid function, $\mathcal{O} = \{(u,i,j)|R_{u,i} = 1, R_{u,j} = 0 \}$ denotes the pairwise training data, and $j$ denotes the sampled item that user $u$ has not interacted with.} 

{By optimizing the BPR loss $\mathcal L_{BPR}$, our proposed \our can model the interactions between users and items.
However, high-order neighbor relations within users (or within items) are also valuable for recommendations. For example, users are more likely to buy the same product as their neighbors. 
Next, we will propose two contrastive learning objectives to capture the potential neighborhood relationships of users and items.}

\subsection{Contrastive Learning with Structural Neighbors} \label{sec:Hop-Contrastive}
Existing graph collaborative filtering models are mainly trained with the observed interactions (\eg user-item pairs), 
while the potential relationships among users or items cannot be explicitly captured by learning from the observed data. 
{In order to fully exploit the advantages of contrastive learning}, we propose to contrast each user~(or item) with his/her structural neighbors whose representations are aggregated through layer propagation of GNN.
Formally, the initial feature or learnable embedding of users/items are denoted by $z^{(0)}$ in the graph collaborative filtering model~\cite{he2020lightgcn}.
And the final output can be seen as a combination of the embeddings within a subgraph that contains multiple neighbors at different hops. Specifically, the $l$-th layer's output $z^{(l)}$ of {the base GNN model} is the weighted sum of $l-$hop structural neighbors of each node, as there is no transformation and self-loop when propagation~\cite{he2020lightgcn}.

\tian{Considering that the interaction graph $\mathcal{G}$ is a bipartite graph, information propagation with GNN-based model for even times on the graph naturally aggregates information of homogeneous structural neighbors which makes it convenient to extract the potential neighbors within users or items. In this way, we can obtain the representations of homogeneous neighborhoods from the even layer (\eg 2, 4, 6) output of {the GNN model}. With these representations, we can efficiently model the relation between users/items and their homogeneous structural neighbors.
Specifically, we treat the embedding of users themself and the embedding of the corresponding output of the even-numbered layer GNN as positive pairs.
Based on InfoNCE~\cite{oord2018representation}, we propose the structure-contrastive learning objective to minimize the distance between them as follows:}
\begin{equation} \label{hop_loss}
\mathcal L_{S}^{U} = \sum_{u\in \mathcal{U}}-\log \frac{\exp((\mathbf {z}_u^{(k)}\cdot \mathbf {z}_u^{(0)}/\tau))}{\sum_{v\in \mathcal{U}} \exp((\mathbf {z}_u^{(k)}\cdot \mathbf {z}_v^{(0)}/\tau))},
\end{equation}    
where $\mathbf {z}_u^{(k)}$ is the normalized output of GNN layer $k$ and $k$ is even number. $\tau$ is the temperature hyper-parameter of softmax. \tian{In a similar way, the structure-contrastive loss of the item side $\mathcal L_{struc}^{item}$ can be obtained as:
\begin{equation} \label{hop_loss_item}
\mathcal L_{S}^{I} = \sum_{i\in \mathcal{I}}-\log \frac{\exp((\mathbf {z}_i^{(k)}\cdot \mathbf {z}_i^{(0)}/\tau))}{\sum_{j\in \mathcal{I}} \exp((\mathbf {z}_i^{(k)}\cdot \mathbf {z}_j^{(0)}/\tau))},
\end{equation}   
And the complete structure-contrastive objective function is the weighted sum of the above two losses:}
\begin{equation} \label{complete_hop_loss}
\mathcal L_{S} = \mathcal L_{S}^{U} + \alpha\mathcal L_{S}^{I} .
\end{equation}
where $\alpha$ is a hyper-parameter \tian{to balance the weight of the two losses in structure-contrastive learning.}

\subsection{Contrastive Learning with Semantic Neighbors} \label{sec:Prototype-Contrastive}

The structure-contrastive loss explicitly excavates the neighbors defined by the interaction graph.
However, the structure-contrastive loss treats the homogeneous neighbors of users/items equally, which inevitably introduces  noise information to contrastive pairs. 
{
To reduce the influence of noise from structural neighbors, 
we consider extending the contrastive pairs by incorporating \emph{semantic neighbors}, which refer to unreachable nodes on the graph but with similar characteristics (item nodes) or preferences (user nodes). 
}

\tian{Inspired by previous works~\cite{lin2021prototypical}, we can identify the semantic neighbors by learning the latent prototype for each user and item.
Based on this idea, we further propose the prototype-contrastive objective to explore potential semantic neighbors and incorporate them into contrastive learning to better capture the semantic characteristics of users and items in collaborative filtering.} In particular, similar users/items tend to  fall in neighboring embedding space,  and the prototypes are the center of clusters that represent a group of semantic neighbors.
{Thus, we apply a clustering algorithm on the embeddings of users and items to obtain the prototypes of {users or items}. 
Since this process cannot be end-to-end optimized, we learn the proposed prototype-contrastive objective with EM algorithm.
Formally, the goal of GNN model is to maximize the following log-likelihood function:}
\begin{equation}\label{re-write_loglikelihood}
\sum_{u \in \mathcal{U}}\log p(\mathbf e_u|\Theta,\textbf{R}) = \sum_{u \in \mathcal{U}}\log \sum_{\mathbf c_i \in C} p(\mathbf e_u,\mathbf c_i|\Theta,\textbf{R}),
\end{equation}
where $\Theta$ is a set of model parameters and $\textbf{R}$ is the interaction matrix. And $c_i$ is the latent prototype of user $u$. Similarly, we can define the optimization objective for items. 

After that, the proposed prototype-contrastive learning objective is to minimize the following function based on InfoNCE~\cite{oord2018representation}:
\begin{equation}\label{eq:prot_loss}
    \mathcal L_{P}^{U} = \sum_{u \in \mathcal{U}} - \log \frac{\exp( \mathbf e_u\cdot \mathbf c_i/ \tau)}{\sum_{\mathbf c_j \in C }\exp(\mathbf e_u\cdot \mathbf c_j/ \tau)} .
\end{equation}
where $c_i$ is the prototype of user $u$ which is got by clustering over all the user embeddings with $K$-means algorithm and there are $k$ clusters over all the users. The objective on the item side is identical:
\begin{equation}\label{eq:prot_loss_item}
    \mathcal L_{P}^{I} = \sum_{i \in \mathcal{I}} - \log \frac{\exp(\mathbf e_i\cdot \mathbf c_j/ \tau)}{\sum_{\mathbf c_t \in C} \exp(\mathbf e_i\cdot \mathbf c_t/ \tau)} .
\end{equation}
where $\mathbf c_j$ is the protype of item $i$. The final prototype-contrastive objective is the weighted sum of user objective and item objective:
\begin{equation}
\mathcal L_{P} = \mathcal L_{P}^{U} + \alpha \mathcal L_{P}^{I}.
\end{equation}
\tian{In this way, we explicitly incorporate the semantic neighbors of users/items into contrastive learning to alleviate the data sparsity.}

\subsection{Optimization} \label{sec:Multi-task}
In this section, we introduce the overall loss and the optimization of the proposed prototype-contrastive objective with EM algorithm.

\paratitle{Overall Training Objective}. As the main target of the collaborative filter is to model the interactions between users and items, we treat the proposed two contrastive learning losses as supplementary and leverage a multi-task learning strategy to jointly train the traditional ranking loss and the proposed contrastive loss.
\begin{equation}
    \mathcal L = \mathcal L_{BPR} + \lambda_1 \mathcal L_{S} + \lambda_2 \mathcal L_{P} + \lambda_3 ||\Theta||_2,
\label{eq:overall_loss}
\end{equation}
{where $\lambda_1$,  $\lambda_2$ and $\lambda_3$ are the hyper-parameters to control the weights of the proposed two objectives and the regularization term, respectively, and 
$\Theta$ denotes the set of GNN model parameters. }

\paratitle{Optimize $\mathcal L_{P} $with EM algorithm.}
As Eq. \eqref{re-write_loglikelihood} is hard to optimize, we obtain its Lower-Bound~(LB) by Jensen's inequality:
\begin{equation}
\begin{aligned}
  LB = \sum_{u \in \mathcal{U}}\sum_{\mathbf c_i \in C} Q(\mathbf c_i|\mathbf e_u)\log\frac{p(\mathbf e_u,\mathbf c_i|\Theta,\textbf{R})}{Q(\mathbf c_i|\mathbf e_u)} ,
\end{aligned}
\end{equation}
where $Q(\mathbf c_i|\mathbf e_u)$ denotes the distribution of latent variable $\mathbf c_i$ when $e_u$ is observed. The target can be redirected to maximize the function over $e_u$ when $Q(\mathbf c_i|\mathbf e_u)$ is estimated. The optimization process is formulated in EM algorithm.

In the E-step, $\mathbf e_u$ is fixed and $Q(\mathbf c_i|\mathbf e_u)$ can be estimated by K-means algorithm over the embeddings of all users $\textbf{E}$. If user $u$ belongs to cluster $i$, then the cluster center $\mathbf c_i$ is the prototype of the user. And the distribution is estimated by a hard indicator $\hat Q(\mathbf c_i|\mathbf e_u) =1$ for $\mathbf c_i$ and $\hat Q(\mathbf c_j|\mathbf e_u) = 0$ for other prototypes $\mathbf c_j$.

In the M-step, the target function can be rewritten with $\hat Q(\bm c_i|\bm e_u)$:
\begin{equation}
   \mathcal L_{P}^{U} = -\sum_{u \in \mathcal{U}}\sum_{\mathbf c_i \in C} \hat Q(\mathbf c_i|\mathbf e_u)\log p(\mathbf e_u,\mathbf c_i|\Theta,\textbf{R}) ,
\end{equation}

we can assume that the distrubution of users is isotropic Gaussian over all the clusters. So the function can be written as:
\begin{equation}
    \mathcal L_{P}^{U} = - \sum_{u \in \mathcal{U}} \log \frac{ \exp(-(\mathbf e_u-\mathbf c_i)^2 / 2\sigma_i^2)}{\sum_{\mathbf c_j \in C} \exp(-(\mathbf e_u-\mathbf c_j)^2 / 2\sigma_j^2)} ,
\end{equation}
As $x_u$ and $c_i$ are normalizated beforehand, then $(\mathbf e_u-\mathbf c_i)^2 = 2-2\mathbf e_u\cdot \mathbf c_i$. Here we make an assumption that each Gussian distribution has the same derivation, which is  written to the temperature hyper-parameter $\tau$. Therefore, the function can be simplified as Eq.~\eqref{eq:prot_loss}.

\subsection{Discussion} \label{sec:Discussion}

\paratitle{Novelty and Differences.}  
For graph collaborative filtering, the construction of neighborhood is more important than other collaborative filtering methods~\cite{wu2020graphsurvey}, since it is based on the graph structure.
To our knowledge, it is the first attempt that leverages both structural and semantic neighbors for graph collaborative filtering.
Although several works~\cite{peng2020graph, li2020prototypical, lin2021prototypical} treat either structural or sematic neighbors as positive contrastive pairs, our work differs from them in several aspects.
For structural neighbors, existing graph contrastive learning methods~\cite{liu2021graph,wu2021sslsurvey, hassani2020contrastive, zhu2020deep, wu2021self} mainly take augmented representations as positive samples, while we take locally aggregated representations as positive samples. Besides, we don't introduce additional graph construction or neighborhood iteration, making NCL more efficient than previous works (\eg SGL~\cite{wu2021self}). Besides, some works~\cite{peng2020graph, wu2021sslsurvey, zhu2020deep} make the contrast between the learned node representations and the input node features, while we make the contrast with representations of homogeneous neighbors, which is more suited to the recommendation task. 

Furthermore, semantic neighbors have seldom been explored in GNNs for recommendation, while semantic neighbors are necessary to be considered for graph collaborative filtering due to the sparse, noisy interaction graphs.
In this work, we apply the prototype learning technique to capture the semantic information, which is different from previous works from computer vision~\cite{li2020prototypical} and graph mining~\cite{jing2021graph,lin2021prototypical,xu2021self}.
First, they aim to learn the inherent hierarchical structure among instances, while we aim to identify nodes with similar preferences/characteristics by capturing underlying associations. 
Second, they model prototypes as clusters of independent instances, while we model prototypes as clusters of highly related users (or items) with similar interaction behaviors.

\paratitle{Time and Space Complexity.} {In the proposed two contrastive learning objectives, assume that we sample $S$ users or items as negative samples.} Then, the time complexity of the proposed method can be roughly estimated as $\mathcal{O} \big(N\cdot (S+K)\cdot d\big)$ where $N$ is the total number of users and items, $K$ is the number of prototypes we defined and $d$ is the dimension embedding vector. When we set $S\ll N$ and $K\ll N$ the total time complexity is approximately linear with the number of users and items. As for the space complexity, the proposed method does not introduce  additional parameters besides the GNN backbone. In particular, our \our save nearly half of space compared to other self-supervised methods~(\eg SGL~\cite{wu2021self}), as we explicitly utilize the relation within users and items instead of explicit data augmentation. In a word, the proposed \our is an efficient and effective contrastive learning paradigm aiming at collaborative filtering tasks.

\section{Experiments}
To verify the effectiveness of the proposed NCL, we conduct extensive experiments and report detailed analysis results. 

\subsection{Experimental Setup}

\subsubsection{Datasets}
To evaluate the performance of the proposed NCL, we use five public datasets to conduct experiments: MovieLens-1M~(ML-1M)~\cite{harper2015movielens}, Yelp\footnote{https://www.yelp.com/dataset}, Amazon Books~\cite{mcauley2015image}, Gowalla~\cite{cho2011friendship} and Alibaba-iFashion~\cite{chen2019pog}. These datasets vary in domains, scale, and density. For Yelp and Amazon Books datasets, we filter out users and items with fewer than 15 interactions to ensure data quality.
The statistics of the datasets are summarized in Table \ref{tab:exp-dataset}.
For each dataset, we randomly select 80\% of interactions as training data and 10\% of interactions as validation data. The remaining 10\% interactions are used for performance comparison. We uniformly sample one negative item for each positive instance to form the training set.

\begin{table}[!t]
	\caption{Statistics of the datasets}
	\label{tab:exp-dataset}
	\begin{tabular}{c *{4}{r}}
		\toprule
		\textbf{Datasets} & \textbf{\#Users} & \textbf{\#Items} & \textbf{\#Interactions} & \textbf{Density}\\
		\midrule
		ML-1M 	& 6,040  & 3,629  & 836,478   & 0.03816  \\
		Yelp 			& 45,478 & 30,709 & 1,777,765 & 0.00127 \\
		Books 	& 58,145 & 58,052 & 2,517,437 & 0.00075 \\
		Gowalla & 29,859 & 40,989 & 1,027,464 & 0.00084 \\
		Alibaba & 300,000 & 81,614 & 1,607,813 & 0.00007  \\
		\bottomrule
	\end{tabular}
\end{table}

\subsubsection{Compared Models}
We compare the proposed method with the following baseline methods.\\
$-$ \textbf{BPRMF}~\cite{rendle2009bpr} optimizes the BPR loss to learn the latent representations for users and items with matrix factorization (MF) framework.\\
$-$ \textbf{NeuMF}~\cite{he2017neural} replaces the dot product in MF model with a multi-layer perceptron to learn the match function of users and items.\\
$-$ \textbf{FISM}~\cite{kabbur2013fism} is an item-based CF model which aggregates the representation of historical interactions as user interest.\\
$-$ \textbf{NGCF}~\cite{wang2019neural} adopts the user-item bipartite graph to incorporate high-order relations and utilizes GNN to enhance CF methods. \\
$-$ \textbf{Multi-GCCF}~\cite{sun2019multi} propagates information among high-order correlation users (and items) besides user-item bipartite graph.\\
$-$ \textbf{DGCF}~\cite{wang2020disentangled} produces disentangled representations for user and item to improve the performance of recommendation.\\
$-$ \textbf{LightGCN}~\cite{he2020lightgcn} simplifies the design of GCN to make it more concise and appropriate for recommendation.\\
$-$ \textbf{SGL}~\cite{wu2021self} introduces self-supervised learning to enhance recommendation. We adopt SGL-ED as the instantiation of SGL.

\subsubsection{Evaluation Metrics}
To evaluate the performance of top-$N$ recommendation, we adopt two widely used metrics Recall@$N$ and NDCG@$N$, where $N$ is set to 10, 20 and 50 for consistency. Following~\cite{he2020lightgcn,wu2021self}, we adopt the full-ranking strategy~\cite{zhao2020revisiting}, which ranks all the candidate items that the user has not interacted with.

\begin{table*}[!ht]
\begin{adjustbox}{max width=2.1\columnwidth}
\begin{threeparttable}
\centering
\caption{Performance Comparison of Different Recommendation Models}
\label{tab:exp-main}
\begin{tabular}{@{}cccccccccccr@{}}
\toprule
Dataset & Metric & \multicolumn{1}{c}{BPRMF} & \multicolumn{1}{c}{NeuMF} & \multicolumn{1}{c}{FISM}  & \multicolumn{1}{c}{NGCF} & MultiGCCF & \multicolumn{1}{c}{DGCF} & LightGCN & SGL & \our & Improv. \\ \midrule \midrule
\multirow{6}{*}{MovieLens-1M} & Recall@10 & \multicolumn{1}{c}{0.1804} & \multicolumn{1}{c}{0.1657} & \multicolumn{1}{c}{0.1887} & \multicolumn{1}{c}{0.1846}  & 0.1830 & \multicolumn{1}{c}{0.1881} & \multicolumn{1}{c}{0.1876} & \multicolumn{1}{c}{\underline{0.1888}} & \textbf{0.2057}$^*$ & +8.95\% \\
 & NDCG@10 & \multicolumn{1}{c}{0.2463} & \multicolumn{1}{c}{0.2295} & \multicolumn{1}{c}{0.2494} & \multicolumn{1}{c}{\underline{0.2528}} & 0.2510 & \multicolumn{1}{c}{0.2520} & 0.2514 & 0.2526 & \textbf{0.2732}$^*$ & +8.07\% \\
 & Recall@20 & \multicolumn{1}{c}{0.2714} & \multicolumn{1}{c}{0.2520} & \multicolumn{1}{c}{0.2798} & \multicolumn{1}{c}{0.2741} & 0.2759 & \multicolumn{1}{c}{0.2779} & 0.2796 & \underline{0.2848}  & \textbf{0.3037}$^*$ & +6.63\% \\
 & NDCG@20 & \multicolumn{1}{c}{0.2569} & \multicolumn{1}{c}{0.2400} & \multicolumn{1}{c}{0.2607} & \multicolumn{1}{c}{0.2614} & 0.2617 & \multicolumn{1}{c}{0.2615} & 0.2620  & \underline{0.2649} & \textbf{0.2843}$^*$ & +7.32\% \\
 & Recall@50 & \multicolumn{1}{c}{0.4300} & \multicolumn{1}{c}{0.4122} & \multicolumn{1}{c}{0.4421} & \multicolumn{1}{c}{0.4341} & 0.4364 & \multicolumn{1}{c}{0.4424} & 0.4469 & \underline{0.4487} & \textbf{0.4686}$^*$ & +4.44\% \\
 & NDCG@50 & \multicolumn{1}{c}{0.3014} & \multicolumn{1}{c}{0.2851} & \multicolumn{1}{c}{0.3078} & \multicolumn{1}{c}{0.3055} & 0.3056 & \multicolumn{1}{c}{0.3078} & 0.3091 & \underline{0.3111}  & \textbf{0.3300}$^*$ & +6.08\% \\ \midrule
\multirow{6}{*}{Yelp} & Recall@10 & 0.0643 & 0.0531 & 0.0714 & 0.0630  & 0.0646 & 0.0723 & 0.0730 & \underline{0.0833}  & \textbf{0.0920}$^*$ & +10.44\%\\
 & NDCG@10 & 0.0458 & 0.0377  & 0.0510 & 0.0446 & 0.0450 & 0.0514 & 0.0520 & \underline{0.0601} & \textbf{0.0678}$^*$ & +12.81\% \\
 & Recall@20 & 0.1043 & 0.0885 & 0.1119 & 0.1026 & 0.1053 & 0.1135 & 0.1163 & \underline{0.1288} & \textbf{0.1377}$^*$ & +6.91\% \\
 & NDCG@20 & 0.0580 & 0.0486 & 0.0636 & 0.0567 & 0.0575 & 0.0641 & 0.0652 & \underline{0.0739} & \textbf{0.0817}$^*$ & +10.55\% \\
 & Recall@50 & 0.1862 & 0.1654 & 0.1963 & 0.1864 & 0.1882 & 0.1989 & 0.2016 & \underline{0.2140} & \textbf{0.2247}$^*$ & +5.00\% \\
 & NDCG@50 & 0.0793 & 0.0685 & 0.0856 & 0.0784 & 0.0790 & 0.0862 & 0.0875 & \underline{0.0964} & \textbf{0.1046}$^*$ & +8.51\% \\ \midrule
\multirow{6}{*}{Amazon-Books} & Recall@10 & 0.0607 & 0.0507 & 0.0721   & 0.0617 & 0.0625 & 0.0737 & 0.0797 & \underline{0.0898} & \textbf{0.0933}$^*$ & +3.90\% \\
 & NDCG@10   & 0.043  &  0.0351 & 0.0504 & 0.0427 & 0.0433 & 0.0521 & 0.0565 & \underline{0.0645} & \textbf{0.0679}$^*$ & +5.27\% \\
 & Recall@20 & 0.0956 &  0.0823 & 0.1099 & 0.0978 & 0.0991 & 0.1128 & 0.1206 & \underline{0.1331} & \textbf{0.1381}$^*$ & +3.76\% \\
 & NDCG@20   & 0.0537 &  0.0447 & 0.0622 & 0.0537 & 0.0545 & 0.064  & 0.0689 & \underline{0.0777} & \textbf{0.0815}$^*$ & +4.89\% \\
 & Recall@50 & 0.1681 &  0.1447 & 0.183  & 0.1699 & 0.1688 & 0.1908 & 0.2012 & \underline{0.2157} & \textbf{0.2175}$^*$ & +0.83\% \\
 & NDCG@50   & 0.0726 &  0.061  & 0.0815 & 0.0725 & 0.0727 & 0.0843 & 0.0899 & \underline{0.0992} & \textbf{0.1024}$^*$ & +3.23\% \\ \midrule
\multirow{6}{*}{Gowalla} & Recall@10 & 0.1158 & 0.1039 & 0.1081 & 0.1192 & 0.1108 & 0.1252 & 0.1362 & \underline{0.1465} & \textbf{0.1500}$^*$ & +2.39\% \\
 & NDCG@10   & 0.0833 & 0.0731 & 0.0755 & 0.0852 & 0.0791 & 0.0902 & 0.0876 & \underline{0.1048} & \textbf{0.1082}$^*$ & +3.24\% \\
 & Recall@20 & 0.1695 & 0.1535 & 0.1620 & 0.1755 & 0.1626 & 0.1829 & 0.1976 & \underline{0.2084} & \textbf{0.2133}$^*$ & +2.35\% \\
 & NDCG@20   & 0.0988 & 0.0873 & 0.0913 & 0.1013 & 0.0940 & 0.1066 & 0.1152 & \underline{0.1225} & \textbf{0.1265}$^*$ & +3.27\% \\
 & Recall@50 & 0.2756 & 0.2510 & 0.2673 & 0.2811 & 0.2631 & 0.2877 & 0.3044 & \underline{0.3197} & \textbf{0.3259}$^*$ & +1.94\% \\
 & NDCG@50   & 0.1450 & 0.1110 & 0.1169 & 0.1270 & 0.1184 & 0.1322 & 0.1414 & \underline{0.1497} & \textbf{0.1542}$^*$ & +3.01\% \\ \midrule
\multirow{6}{*}{Alibaba-iFashion} & Recall@10 & 0.303 & 0.182 & 0.0357 & 0.0382 & 0.0401 & 0.0447 & 0.0457 & \underline{0.0461} & \textbf{0.0477}$^*$ & +3.47\% \\
 & NDCG@10   & 0.0161 & 0.0092 & 0.0190 & 0.0198 & 0.0207 & 0.0241 & 0.0246 & \underline{0.0248} & \textbf{0.0259}$^*$ & +4.44\% \\
 & Recall@20 & 0.0467 & 0.0302 & 0.0553 & 0.0615 & 0.0634 & 0.0677 & 0.0692 & \underline{0.0692} & \textbf{0.0713}$^*$ & +3.03\% \\
 & NDCG@20   & 0.0203 & 0.0123 & 0.0239 & 0.0257 & 0.0266 & 0.0299 & 0.0246 & \underline{0.0307} & \textbf{0.0319}$^*$ & +3.01\% \\
 & Recall@50 & 0.0799 & 0.0576 & 0.0943 & 0.1081 & 0.1107 & 0.1120 & \underline{0.1144} & 0.1141 & \textbf{0.1165}$^*$ & +1.84\%\\
 & NDCG@50   & 0.0269 & 0.0177 & 0.0317 & 0.0349 & 0.0360 & 0.0387 & 0.0396 & \underline{0.0396} & \textbf{0.0409}$^*$ & +3.28\%\\ \bottomrule
\end{tabular}
\begin{tablenotes}
      \item The best result is \textbf{bolded} and the runner-up is \underline{underlined}. $^*$ indicates the statistical significance for $p$ < 0.01 compared to the best baseline.
\end{tablenotes}
\end{threeparttable}
\end{adjustbox}
\end{table*}

\subsubsection{Implementation Details}
{We implement the proposed model and all the baselines with \texttt{RecBole}\footnote{\href{https://github.com/RUCAIBox/RecBole}{https://github.com/RUCAIBox/RecBole}}~\cite{zhao2021recbole}, which is a unified open-source framework to develop and reproduce recommendation algorithms.}
To ensure a fair comparison, we optimize all the methods with Adam optimizer and carefully search the hyper-parameters of all the baselines. The batch size is set to 4,096 and all the parameters are initialized by the default Xavier distribution. The embedding size is set to 64. We adopt early stopping with the patience of 10 epoch to prevent overfitting, and NDCG@10 is set as the indicator. We tune the hyper-parameters $\lambda_1$ and $\lambda_2$ in [1e-10,1e-6], $\tau$ in [0.01,1] and $k$ in [5,10000].

\subsection{Overall Performance}

Table \ref{tab:exp-main} shows the performance comparison of the proposed \our and other baseline methods on five datasets. From the table, we find several observations:

(1) Compared to the traditional methods, such as BPRMF, GNN-based methods outperform as they encode the high-order information of bipartite graphs into representations. {Among all the  graph collaborative filtering baseline models,} LightGCN performs best in most datasets, which shows  the effectiveness and robustness of the simplified architecture~\cite{he2020lightgcn}. Surprisingly, Multi-GCCF performs worse than NGCF on ML-1M, probably because the projection graphs built directly from user-item graphs are  so dense that the neighborhoods of different users or items on the projection graphs are overlapping and indistinguishable.
Besides, the disentangled representation learning method DGCF is worse than LightGCN, especially on the sparse dataset. We speculate that the dimension of disentangled representation may be too low to carry adequate characteristics as we astrict the overall dimension.
In addition, FISM performs better than NGCF on three datasets (ML-1M, Yelp, and AmazonBooks), {indicating that 
a heavy GNN architecture is likely to overfit over sparse user-item interaction data. }

(2) 
For the self-supervised method, SGL~\cite{wu2021self} consistently outperforms other supervised methods on five datasets, {which shows the effectiveness of contrastive learning for improving the recommendation performance.} 
However, SGL contrasts the representations derived from the original graph with an augmented graph, which neglects other potential relations (\eg user similarity) in recommender systems. 

(3) 
Finally, we can see that the proposed \our consistently performs better than baselines.
{This advantage is brought by the neighborhood-enriched contrastive learning objectives.}
{Besides, the improvement at smaller positions (\eg top $10$ ranks) is greater than that at larger positions (\eg top $50$ ranks), indicating that \our tends to rank the relevant items higher, which is significative in real-world recommendation scenario.}
{In addition, our method yields more improvement on small datasets, such as ML-1M and Yelp datasets.}
We speculate that a possible reason is that the interaction data of those datasets are more sparse, and there are not sufficient neighbors to construct the contrastive pairs. 

\subsection{Further Analysis of \our} \label{sec:further}
In this section, we further perform a series of detailed analysis on the proposed NCL to confirm its effectiveness.
Due to the limited space, we only report the results on ML-1M and Yelp datasets, and the observations are similar on other datasets.

\subsubsection{Ablation Study of NCL}
\begin{figure}[!t]
	{
		\begin{minipage}[t]{0.47\linewidth}
			\centering
			\includegraphics[width=1\textwidth]{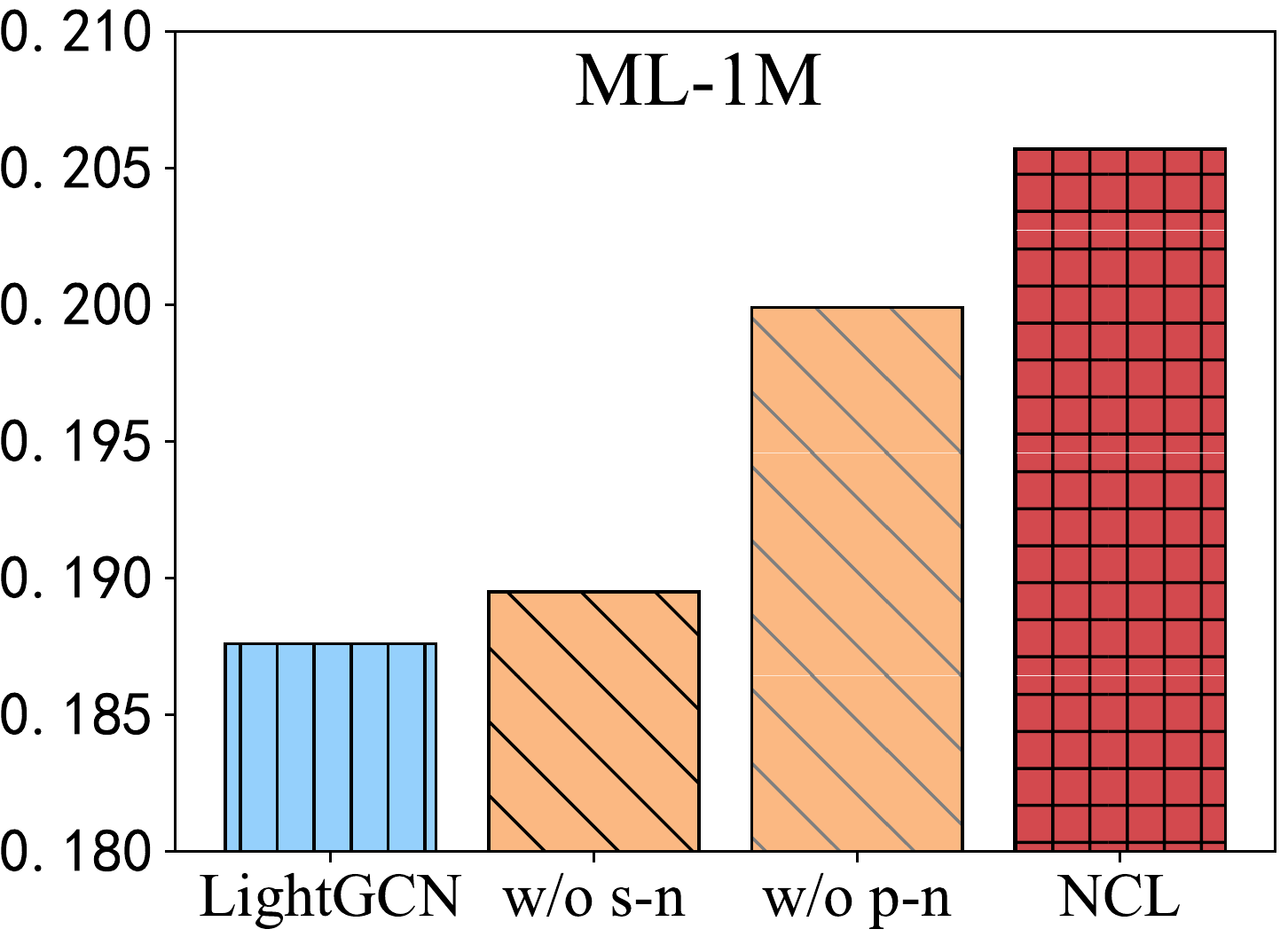}
			 \label{fig:exp_para_alpha_1m}
		\end{minipage}
		\begin{minipage}[t]{0.46\linewidth}
			\centering
			\includegraphics[width=1\textwidth]{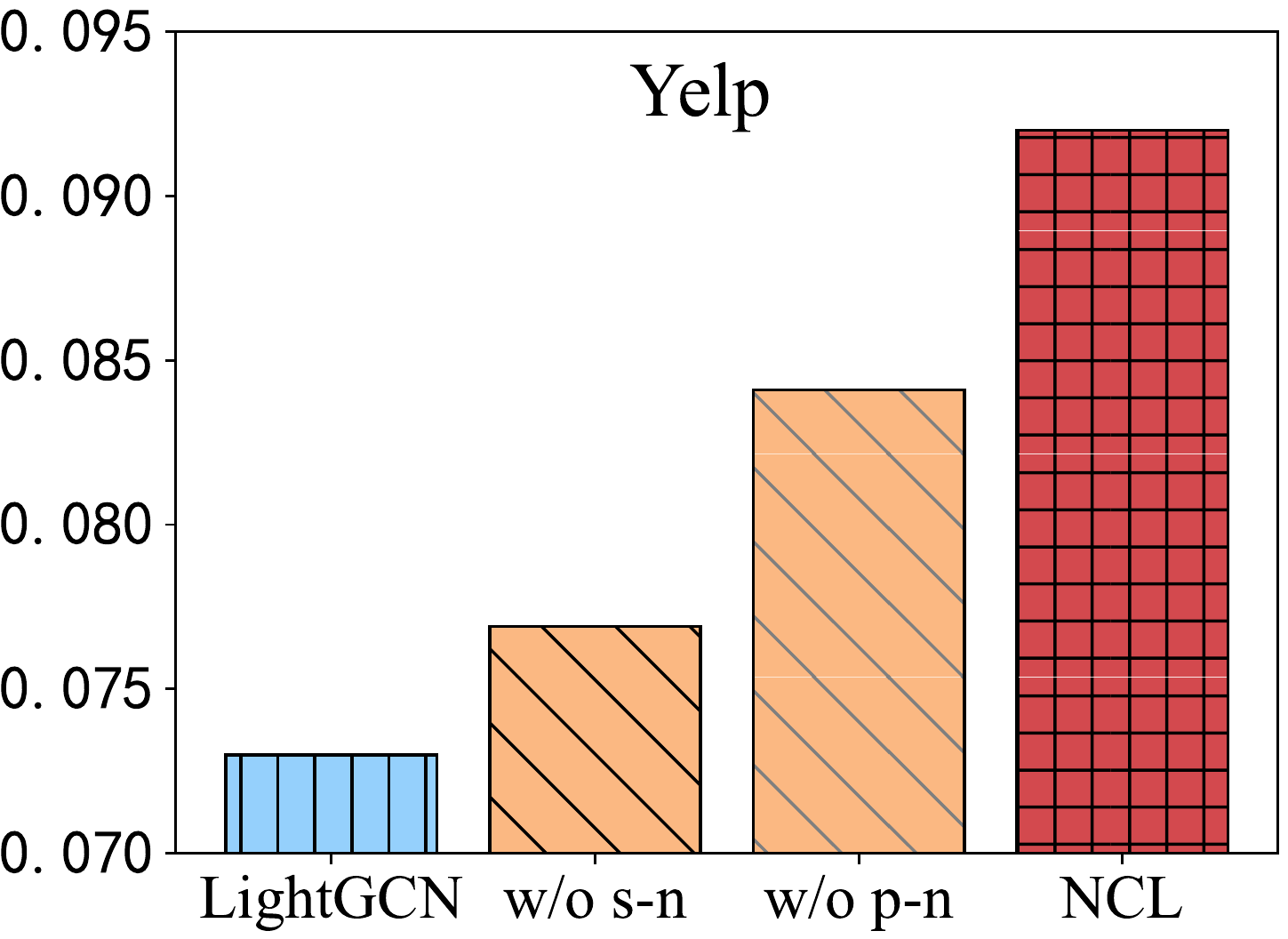}
			\label{fig:exp_para_alpha_yelp}
		\end{minipage}
	}
	\vspace{-0.2in}
	\caption{Performance of \our on two datasets without structural neighbors and semantic neighbors~(Recall@10).} \label{fig:ablation}
\end{figure}

Our proposed approach \our leverages the potential neighbors in two aspects. 
To verify the effectiveness of each kind of neighbor, we conduct the ablation study to analyze their contribution. The results are reported in Figure~\ref{fig:ablation}, where "w/o s-n" and "w/o p-n" denote the variants by removing structural neighbors and semantic neighbors, respectively.
From this figure, we can observe that removing each of the relations leads to the performance decrease while the two variants are both perform better than the baseline LightGCN. It indicates that explicitly modeling both kinds of relations will benefit the performance in graph collaborative filtering. Besides, these two relations complement each other and improve the performance in different  aspects.

\subsubsection{{Impact of Data Sparsity Levels.}}
To further verify the proposed \our can alleviate the sparsity of interaction data, we evaluate the performance of \our on users with different sparsity levels in this part.
\begin{figure}[t]
	{
		\begin{minipage}[t]{0.45\linewidth}
			\centering
			\includegraphics[width=1\textwidth]{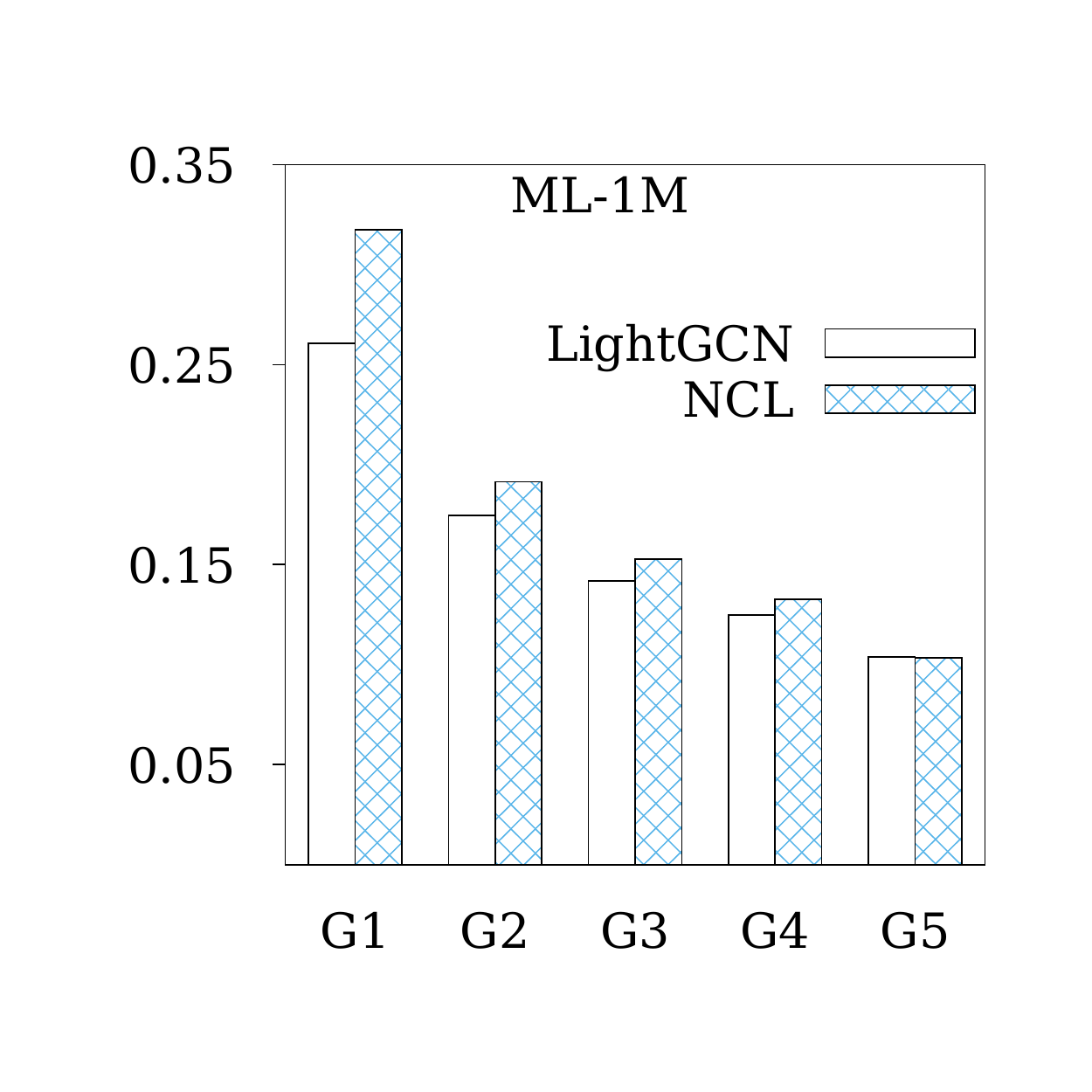}
		\end{minipage}
		\hspace{0.2in}
		\begin{minipage}[t]{0.45\linewidth}
			\centering
			\includegraphics[width=1\textwidth]{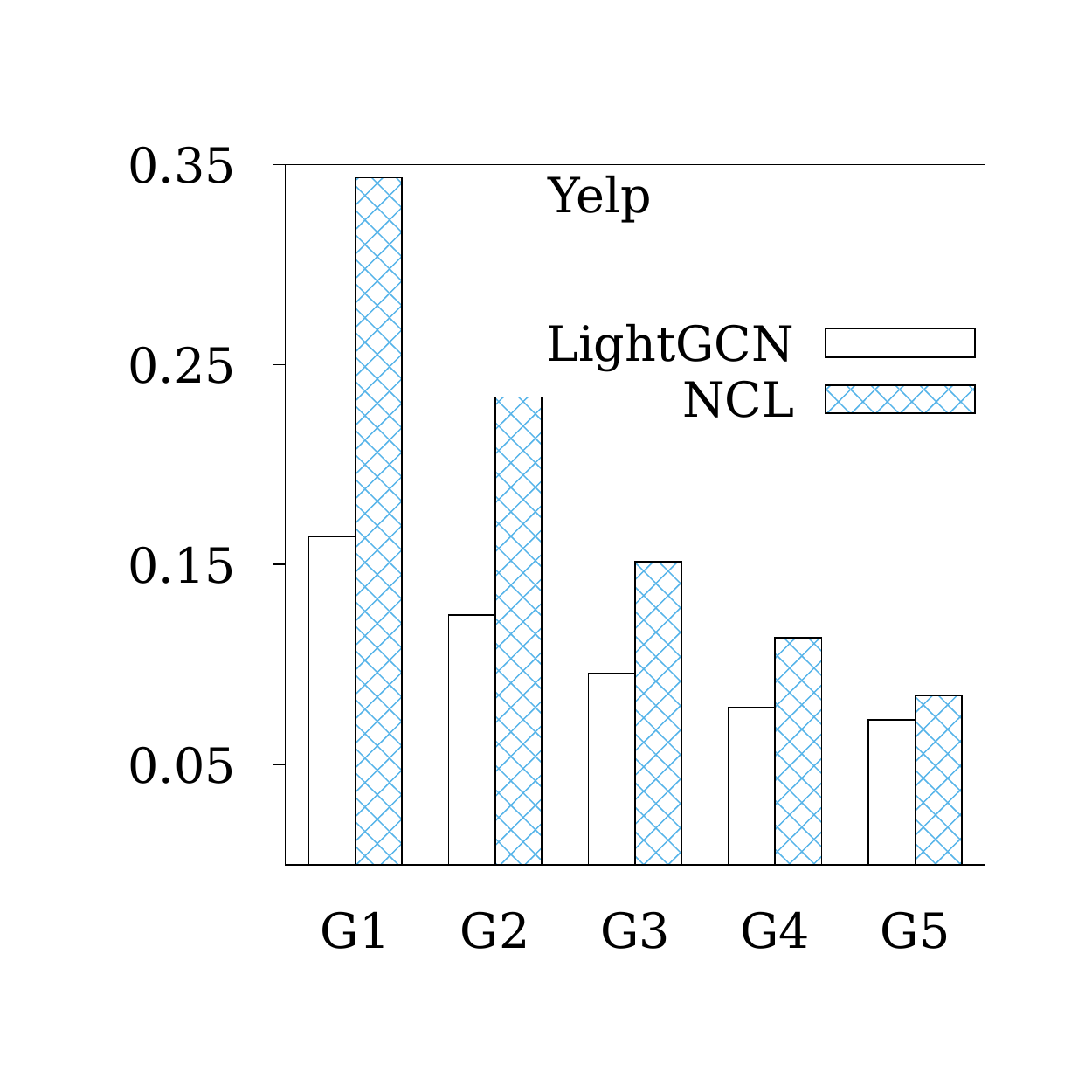}
		\end{minipage}
	}
	\vspace{-0.1in}
	\caption{Performance analysis for different sparsity-level users~(Recall@10). $G1$ denotes the group of users with the lowest average number of interactions. } \label{fig: Long-tail}
\end{figure}
Concretely, we split all the users into five groups based on their interaction number, while keeping the total number of interactions in each group constant.
Then, we compare the recommendation performance of \our and LightGCN on these five groups of users and report the results in Figure~\ref{fig: Long-tail}.
From this figure, we can find that the performance of \our is consistently better than LightGCN.
Meanwhile, as the number of interactions decreases, the performance gain brought by \our increases. 
This implies that \our can perform high-quality recommendation with sparse interaction data, benefited by the proposed neighborhood modeling techniques.

\subsubsection{Effect of Structural Neighbors} In \our, the structural neighbors correspond to different layers of GNN. To investigate the impact of different structural neighbors, we select the nodes in one-, two-, and three-hop as the structural neighbors and test the effectiveness when incorporating them with contrastive learning. 

The results are shown in Table \ref{tab:div}. We can find that the three variants of \our all perform similar or better than LightGCN, which further indicates the effectiveness of the proposed hop-contrastive strategy. Specifically, the results of the first even layer are the best among these variants. 
{This accords with the intuition that users or items should be more similar to their direct neighbors than indirect neighbors. Besides, in our experiments,  one-hop neighbors seem to be sufficient for \our, making a good trade-off between effectiveness and efficiency. }

\begin{table}[h]
\caption{Performance comparison w.r.t. different hop of structural neighbors.}
\small
\label{tab:div}
\begin{tabular}{c cccccc}
\toprule
\multirow{2}{*}{Hop} &\multicolumn{2}{c}{MovieLens-1M} & \multicolumn{2}{c}{Yelp} \\
            & Recall@10 & NDCG@10 & Recall@10 & NDCG@10 \\ \midrule 
w/o s-n    & 0.1876    & 0.2514  & 0.0730    & 0.0520  \\
1           & 0.2057    & 0.2732  & 0.0920    & 0.0678  \\
2           & 0.1838    & 0.2516  & 0.0837    & 0.0602  \\
3           & 0.1839    & 0.2507  & 0.0787    & 0.0557  \\
\bottomrule
\end{tabular}
\end{table}

\subsubsection{Impact of the Coefficient $\alpha$.}
\begin{figure}[t]
	{
		\begin{minipage}[t]{0.325\linewidth}
			\centering
			\includegraphics[width=1\textwidth]{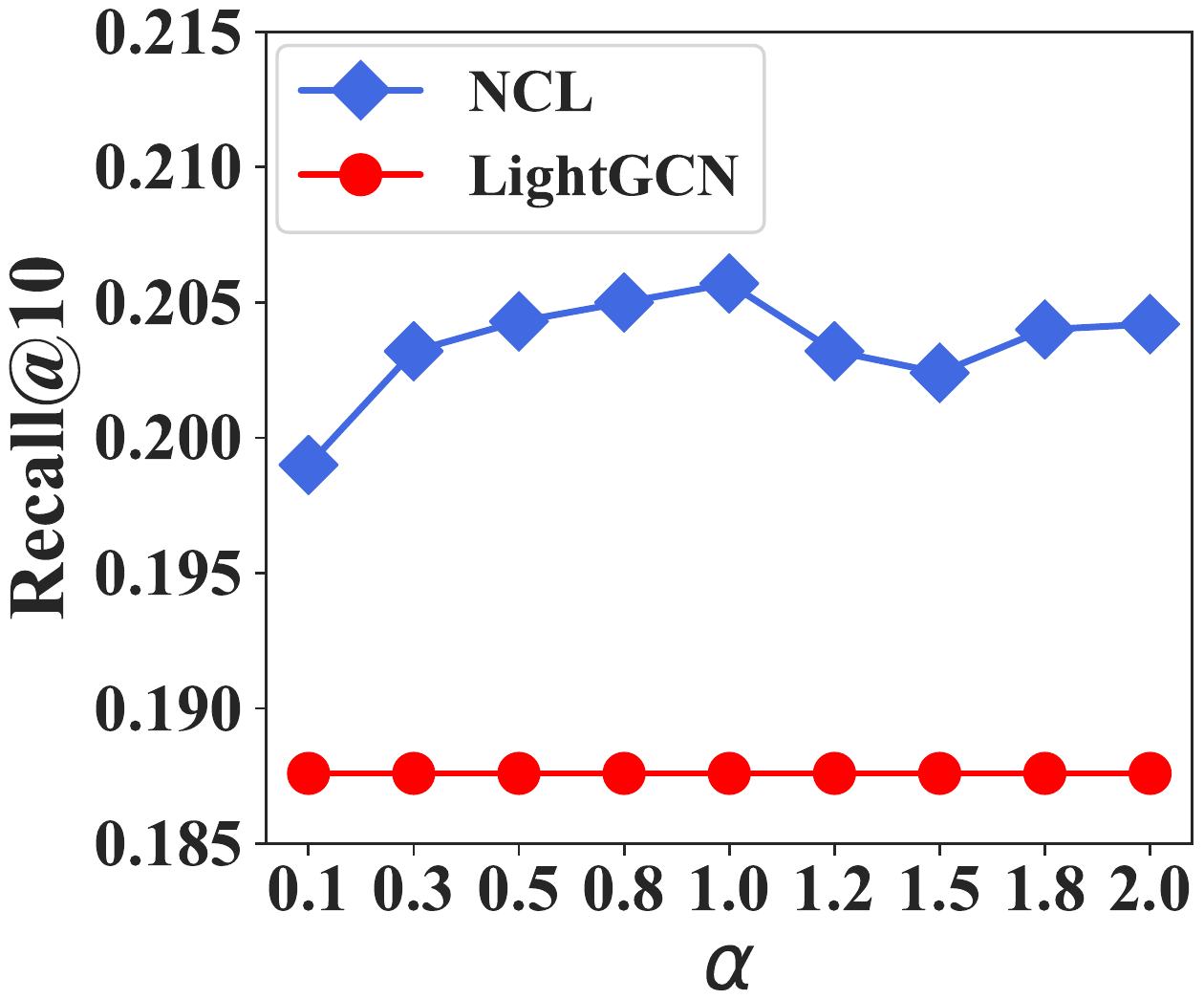}
			\includegraphics[width=1\textwidth]{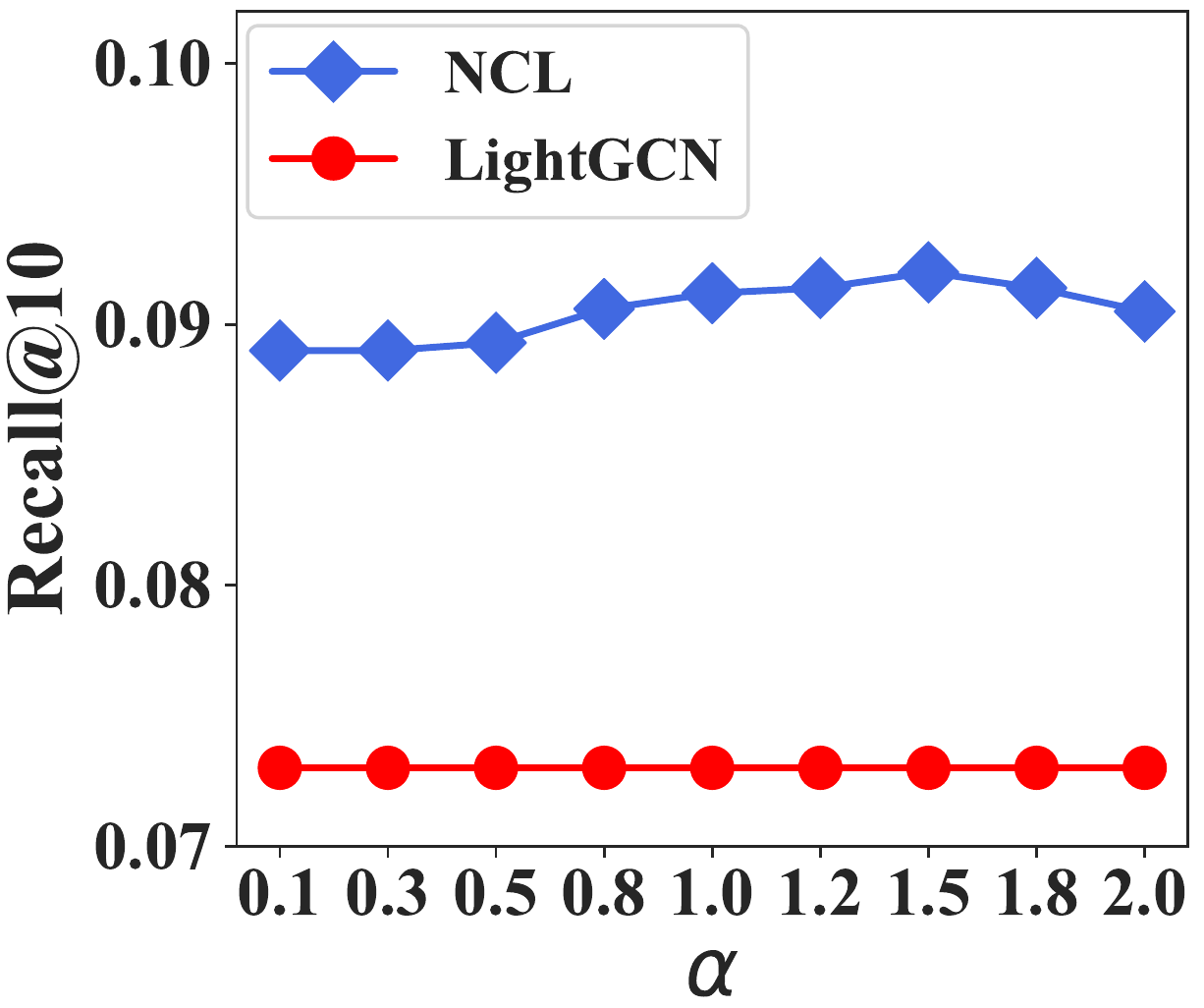}
			\subcaption{} \label{fig:exp_para_alpha}
		\end{minipage}
		\begin{minipage}[t]{0.325\linewidth}
			\centering
			\includegraphics[width=1\textwidth]{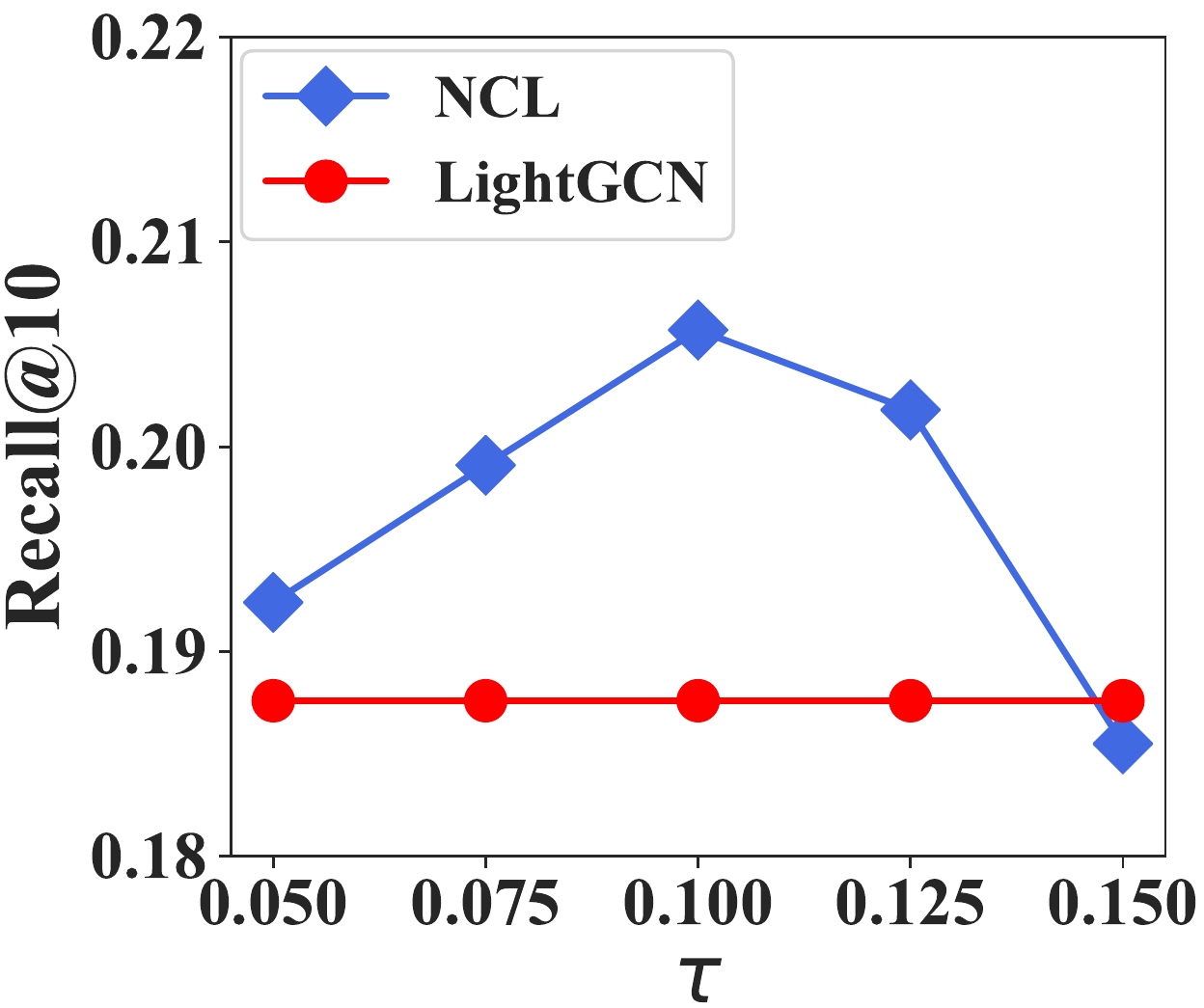}
			\includegraphics[width=1\textwidth]{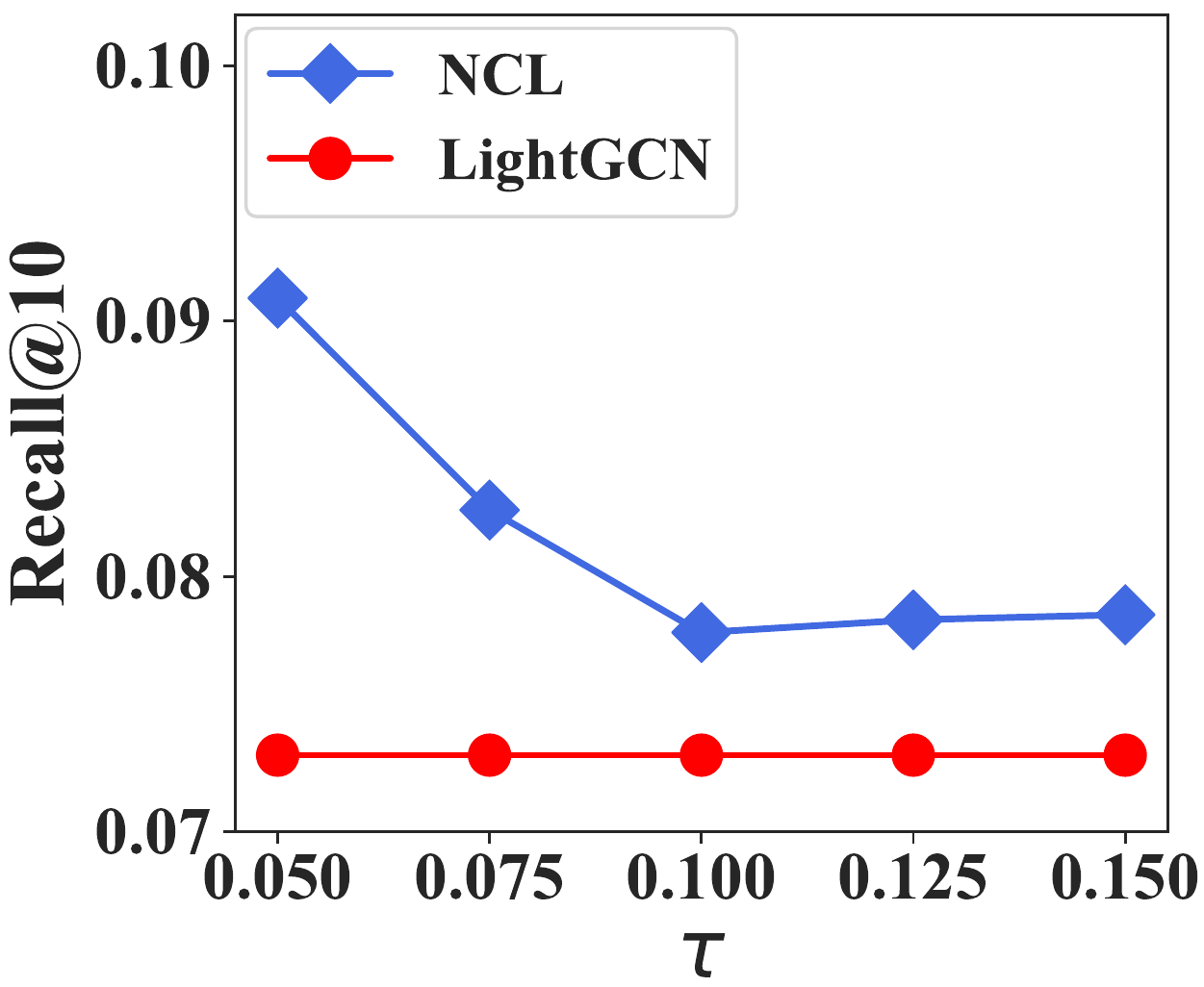}
			\subcaption{} \label{fig:exp_para_tau}
		\end{minipage}
		\begin{minipage}[t]{0.325\linewidth}
			\centering
			\includegraphics[width=1\textwidth]{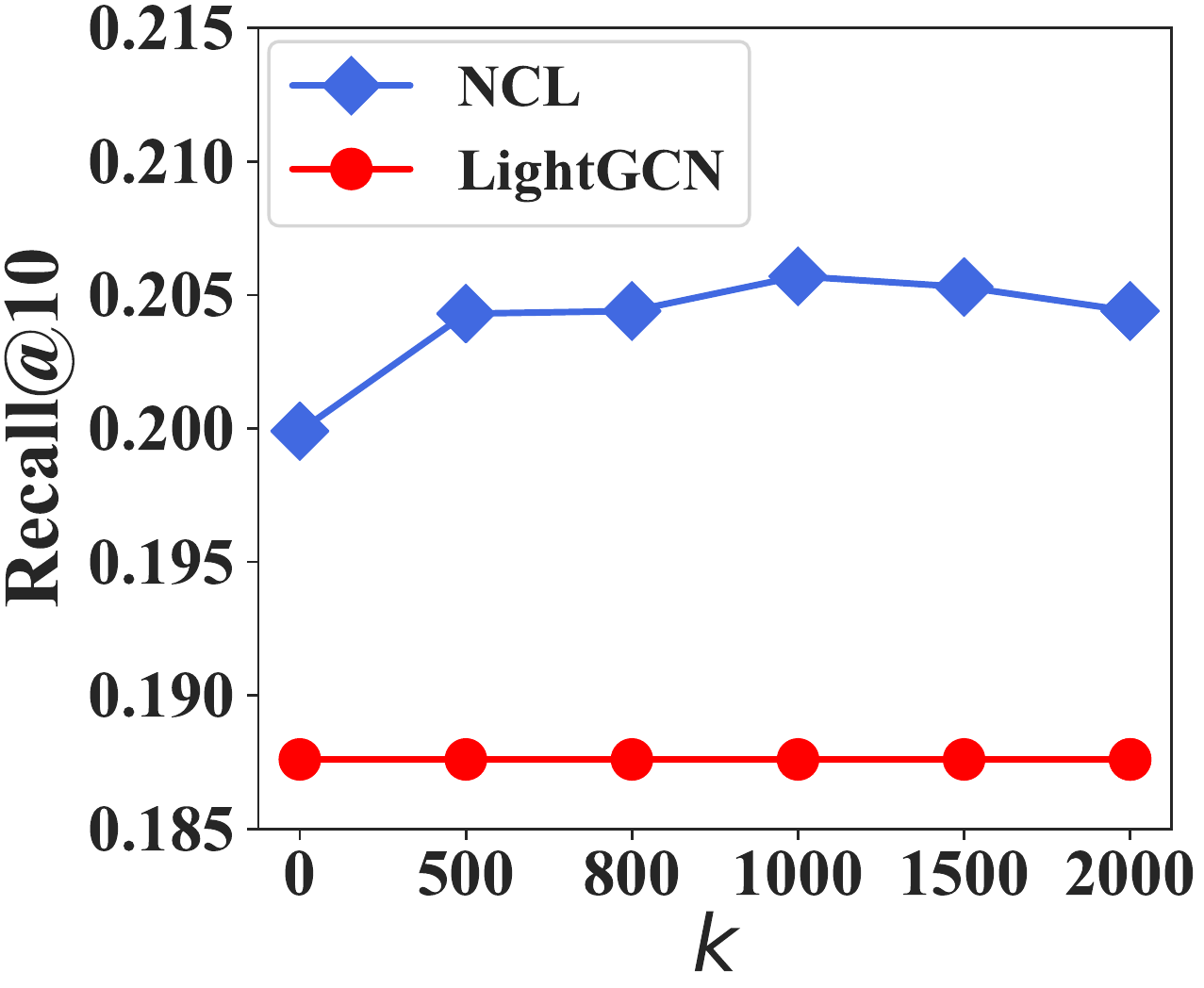}
			\includegraphics[width=1\textwidth]{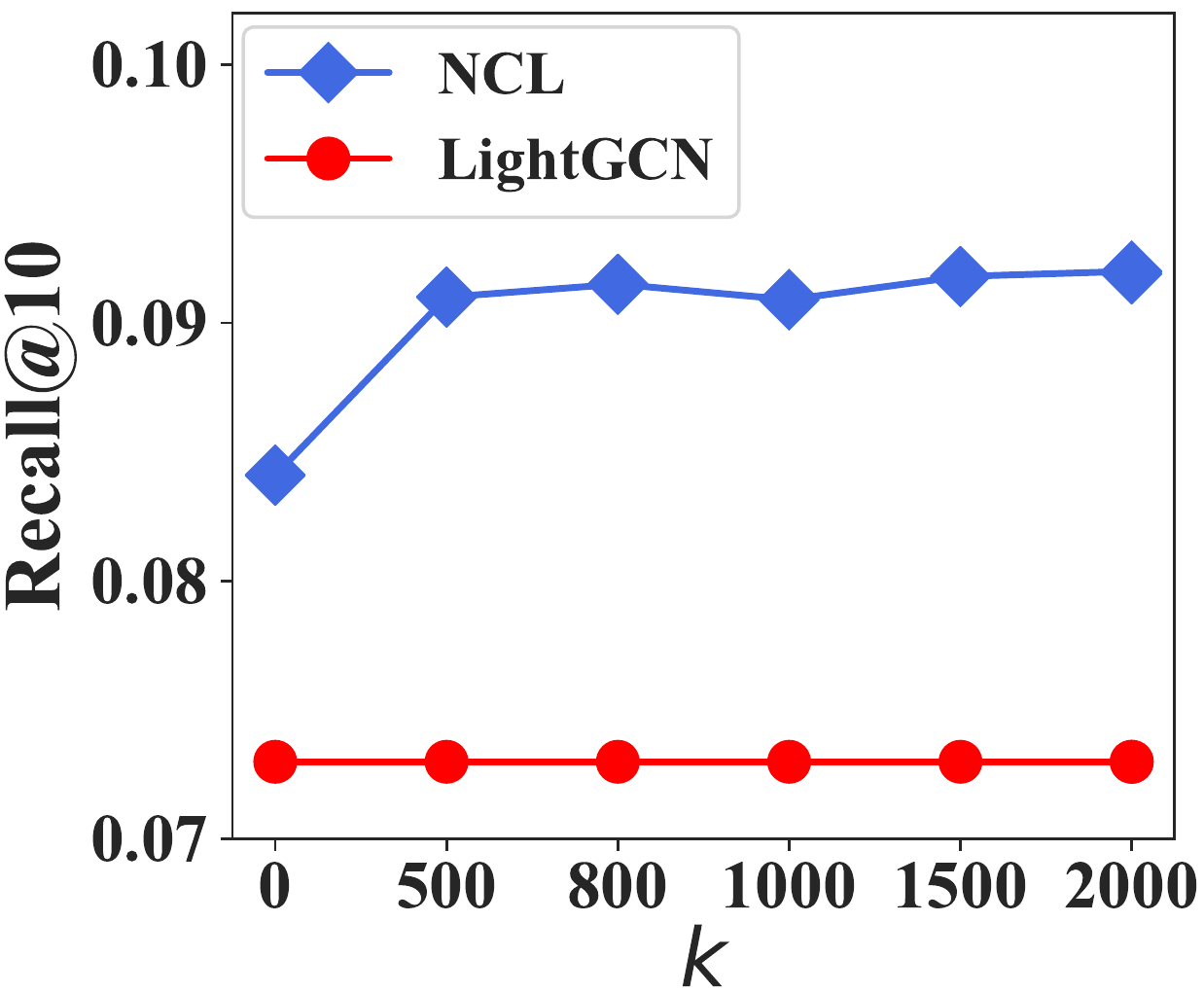}
			\subcaption{} \label{fig:exp_para_k}
		\end{minipage}
	}
	\vspace{-0.15in}
	\caption{Performance comparison w.r.t. different $\alpha$, $\tau$ and $k$. The top shows the Recall@10 results on MovieLens-1M and the bottom shows the results on Yelp.}
\end{figure}

In the structure-contrastive loss defined in Eq.~\eqref{complete_hop_loss}, the coefficient $\alpha$ can balance the two losses for structural neighborhood modeling.
To analyze the influence of $\alpha$, we vary $\alpha$ in the range of 0.1 to 2 and report the results in Figure~\ref{fig:exp_para_alpha}.
It shows that an appropriate $\alpha$ can effectively improve the  performance of NCL.
Specifically, when the hyper-parameter $\alpha$ is set to around 1, the performance is better on both datasets, indicating that the high-order similarities of both users and goods are valuable.
In addition, with different $\alpha$, the performance of \our is consistently better than that of LightGCN, which indicates that \our is robust to parameter $\alpha$.

\subsubsection{Impact of the Temperature $\tau$.}
As in previous works mentioned~\cite{chen2020simple,you2020graph}, the temperature $\tau$ defined in Eq.(\ref{hop_loss}) and  Eq.(\ref{eq:prot_loss}) plays an important role in contrastive learning. To analyze the impact of temperature on NCL, we vary $\tau$ in the range of 0.05 to 0.15 and show the results in Figure~5(b). 
We can observe that  a too large value of $\tau$ will cause poor performance, which is consistent with the experimental results reported in \cite{you2020graph}. 
{In addition, the suitable temperature corresponding to Yelp dataset is smaller, which indicates that the temperature of NCL should be smaller on more sparse datasets.}
Generally, a temperature  in the range of $[0.05, 0.1]$ can lead to good recommendation performance.

\subsubsection{Impact of the Prototype Number $k$.}
To study the effect of prototype-contrastive objective, we set the number of prototypes $k$ from hundreds to thousands and remove it by setting $k$ as zero. The results are reported in Figure~5(c).
As shown in Figure~5(c), NCL with different $k$ consistently outperforms the baseline and the best result is achieved when $k$ is around 1000. 
{It indicates that a large number of prototypes can better mitigate the noise introduced by structural neighbors. 
When we set $k$ as zero, the performance decreases significantly, which shows that semantic neighbors are very useful to improve the recommendation performance.}

\subsubsection{Applying \our on Other GNN Backbones.}
{As the proposed \our architecture is model agnostic, we further test its performance with other GNN architectures. The results are reported in Table \ref{tab:ablation}.
From this table, we can observe that the proposed method can consistently improve the performance of NGCF, DGCF, and LightGCN, which further verifies the effectiveness of the proposed method.} 
Besides, the improvement on NGCF and DGCF is not as remarkable as the improvement on LightGCN. A possible reason is that LightGCN removes the parameter and non-linear activation in layer propagation which ensures the output of different layers in the same representation space for structural neighborhood modeling.

\begin{table}[t]
\centering
\small
\caption{Performance comparison w.r.t.  different GNN backbones.}
\label{tab:ablation}
\begin{tabular}{ccccc}
\toprule
\multicolumn{1}{c}{\multirow{2}{*}{Method}} & \multicolumn{2}{c}{MovieLens-1M}                            & \multicolumn{2}{c}{Yelp}                                    \\
\multicolumn{1}{c}{}                        & \multicolumn{1}{c}{Recall@10} & \multicolumn{1}{c}{NDCG@10} & \multicolumn{1}{c}{Recall@10} & \multicolumn{1}{c}{NDCG@10} \\ \midrule
NGCF                                        & \multicolumn{1}{c}{0.1846}          &  \multicolumn{1}{c}{0.2528}        & \multicolumn{1}{c}{0.0630}          & \multicolumn{1}{c}{0.0446}        \\
\textbf{+NCL}    & \textbf{0.1852} & \textbf{0.2542} & \textbf{0.0663} & \textbf{0.0465} \\ \midrule
DGCF     & 0.1853 & 0.2500 & 0.0723 & 0.0514 \\
\textbf{+NCL}    &  \textbf{0.1877} &  \textbf{0.2522} & \textbf{0.0739}  & \textbf{0.0528}  \\ \midrule
LightGCN & 0.1888 & 0.2526 & 0.0833 & 0.0601 \\
\textbf{+NCL}    & \textbf{0.2057} & \textbf{0.2732} & \textbf{0.0920} & \textbf{0.0678} \\ \bottomrule
\end{tabular}
\end{table}

\begin{figure}[t]
    \begin{minipage}[t]{0.22\linewidth}
		\centering
		\includegraphics[width=1\textwidth]{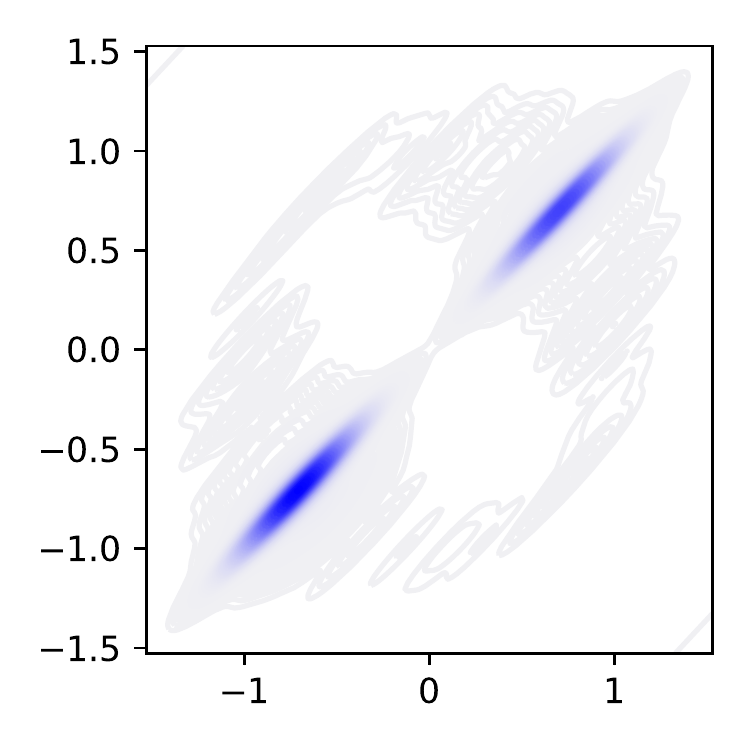}
		\subcaption{LightGCN} \label{fig:kde_ml1m_lightgcn}
	\end{minipage}
	\begin{minipage}[t]{0.22\linewidth}
		\centering
		\includegraphics[width=1\textwidth]{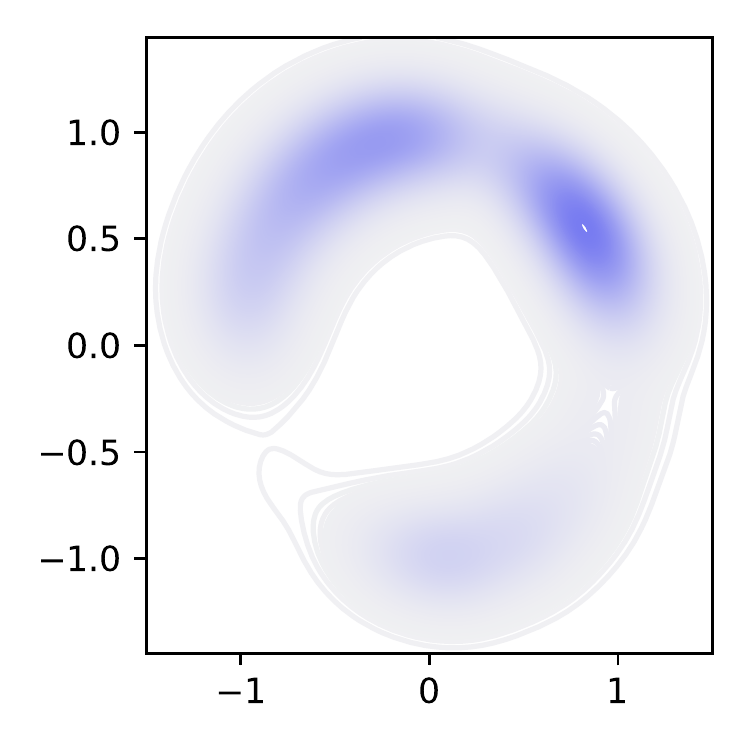}
		\subcaption{NCL} \label{fig:kde_ml1m_ncl}
	\end{minipage}
    \begin{minipage}[t]{0.014\linewidth}
        \centering
        \includegraphics[width=1\textwidth]{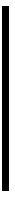}
    \end{minipage}
	\begin{minipage}[t]{0.22\linewidth}
		\centering
		\includegraphics[width=1\textwidth]{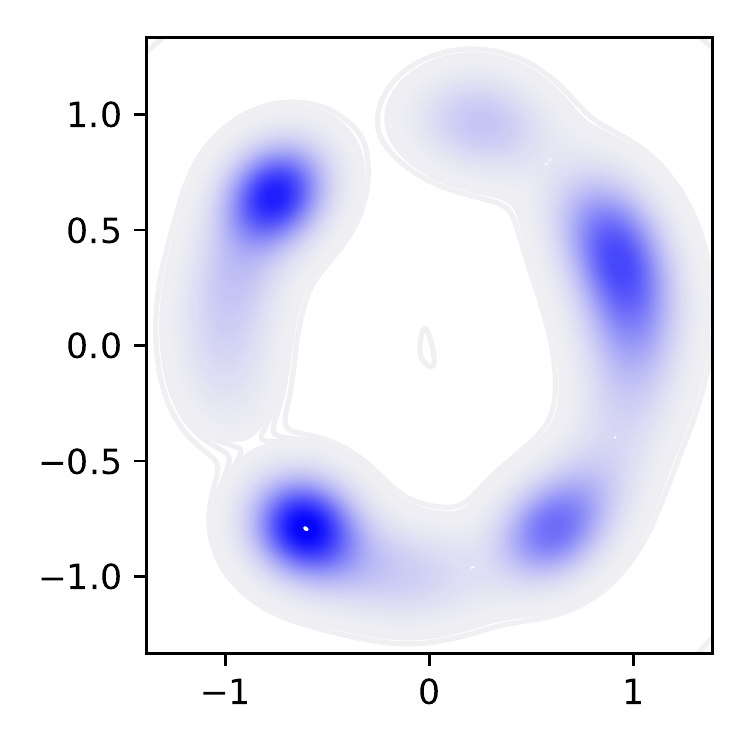}
		\subcaption{LightGCN} \label{fig:kde_yelp_lightgcn}
	\end{minipage}
	\begin{minipage}[t]{0.22\linewidth}
		\centering
		\includegraphics[width=1\textwidth]{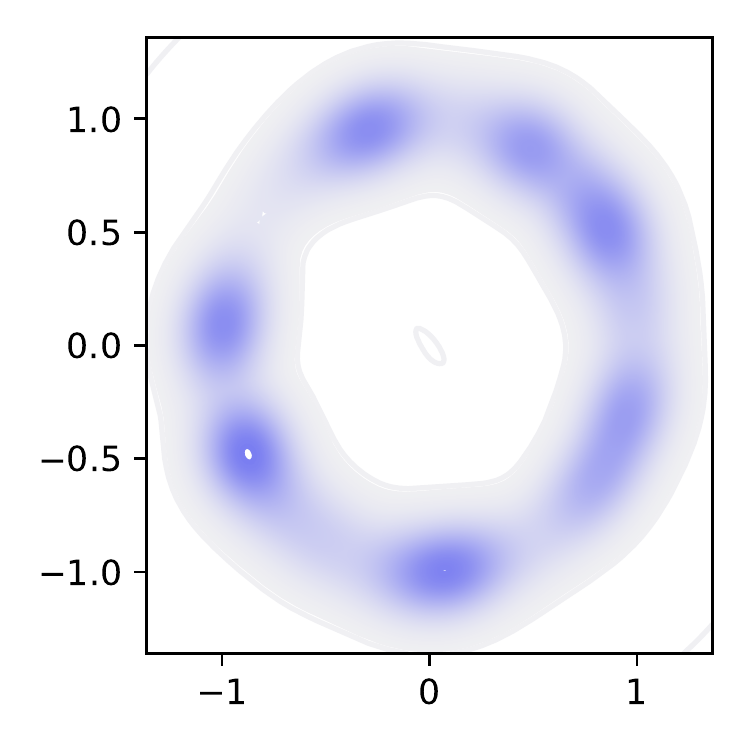}
		\subcaption{NCL} \label{fig:kde_yelp_ncl}
	\end{minipage}
    \vspace{-0.1in}
    \caption{Visualization of item embeddings. Items from ML-1M and Yelp are illustrated in \textbf{(a)}, \textbf{(b)} and \textbf{(c)}, \textbf{(d)}, respectively.}
    \label{fig:kde}
\end{figure}

\subsubsection{Visualizing the Distribution of Representations}
A key contribution of the proposed NCL is to integrate two  kinds of neighborhood relations in the contrastive tasks for graph collaborative filtering. To better understand the benefits brought by NCL, we visualize the learned embeddings in Figure~\ref{fig:kde} to show how the proposed approach affects representation learning.
We plot item embedding distributions with Gaussian kernel density estimation (KDE) in two-dimensional space.
We can see that, embeddings learned by LightGCN fall into several coherent clusters,
while those representations learned by NCL clearly exhibit a more uniform distribution.
{We speculate that a more uniform distribution of embeddings endows a better capacity to model the diverse user preferences or item characteristics.}
As shown in previous studies~\cite{wang2020understanding}, there exists strong correlation between contrastive learning and  \emph{uniformity} of the learned representations, where it prefers a feature distribution that preserves maximal information about representations.
\section{Related work}
In this section, we briefly review the related works in two aspects, namely  graph-based collaborative filtering and contrastive learning.

\paratitle{Graph-based collaborative filtering}.
Different from traditional CF methods, such as matrix factorization-based methods ~\cite{rendle2009bpr,koren2009matrix} and auto-encoder-based methods~\cite{liang2018variational,strub2015collaborative}, graph-based collaborative filtering organize interaction data into an interaction graph and learn meaningful node representations from the graph structure information. Early studies~\cite{gori2007itemrank,baluja2008video} extract the structure information through random walks in the graph. Next, Graph Neural Networks~(GNN) are adopted on collaborative filtering~\cite{he2020lightgcn,wang2019neural,wang2020disentangled,ying2018graph}.
For instance, NGCF~\cite{wang2019neural} and LightGCN~\cite{he2020lightgcn} leverage the high-order relations on the interaction graph to enhance the recommendation performance. 
{Besides, some studies~\cite{sun2019multi} further propose to construct more interaction graphs to capture more 
rich association relations among users and items.}
Despite the effectiveness, they don't explicilty address the data sparsity issue.
More recently, self-supervised learning is introduced into graph collaborative filtering to improve the generalization of recommendation. For example, SGL~\cite{wu2021self} devise random data argumentation operator and construct the contrastive objective to improve the accuracy and robustness of GCNs for recommendation. 
However, most of the graph-based methods only focus on interaction records but neglect the potential neighbor relations among users or items.

\paratitle{Contrastive learning}.
Since the success of contrastive learning in CV~\cite{chen2020simple}, contrastive learning has been widely applied on NLP~\cite{giorgi2020declutr}, graph data mining~\cite{liu2021graph,wu2021sslsurvey} and recommender systems~\cite{xia2020self,tang2021multi}. As for graph contrastive learning, existing studies can be categorized into \emph{node-level} contrastive learning~\cite{velivckovic2018deep,zhu2021graph} and \emph{graph-level} contrastive learning~\cite{you2020graph,sun2019infograph}. 
For instance, GRACE~\cite{zhu2020deep} proposes a framework for node-level graph contrastive  learning, and performs corruption by removing edges and masking node features. MVGRL~\cite{hassani2020contrastive} transforms graphs by graph diffusion, which considers the augmentations in both feature and structure spaces on graphs.
Besides,inspired by the pioneer study in computer vision~\cite{li2020prototypical}, several methods~\cite{jing2021graph,lin2021prototypical,xu2021self} are proposed to adopt prototypical contrastive learning to capture the semantic information in graphs.
Related to our work, several studies also apply contrastive learning to recommendation, such as SGL~\cite{wu2021self}. 
However, existing methods construct the contrastive pairs by random sampling, and do not fully consider the relations among users~(or items) in recommendation scenario.
In this paper, we propose to explicilty model these potential neighbor relations via contrastive learning.

\section{Conclusion And Future Work}
In this work, we propose a novel contrastive learning paradigm, named \textbf{N}eighborhood-enriched \textbf{C}ontrastive \textbf{L}earning~(\textbf{NCL}), to explicitly capture potential node relatedness into contrastive learning for graph collaborative filtering. 
We consider the neighbors of users~(or items) from the two aspects of graph structure and semantic space, respectively.
Firstly, to leverage structural neighbors  on the interaction graph, we develop a novel structure-contrastive objective that can be combined with GNN-based collaborative filtering methods. 
{Secondly, to leverage semantic neighbors, we derive the prototypes of users/items by clustering the embeddings and incorporating the semantic neighbors into the prototype-contrastive objective.}
Extensive experiments on five public datasets demonstrate the effectiveness of the proposed \our.

{As future work, we will extend our framework to other 
recommendation tasks, such as sequential recommendation. Besides, we will also consider developing  a more unified formulation for leveraging and utilizing  different kinds of neighbors. 
}

\begin{acks}
This work was partially supported by the National Natural Science Foundation of China under Grant No. 61872369 and 61832017,
Beijing Outstanding Young Scientist Program under Grant No. BJJWZYJH012019100020098.
This work is supported by Beijing Academy of Artificial Intelligence(BAAI).
Xin Zhao is the corresponding author.
\end{acks}


\bibliographystyle{ACM-Reference-Format}
\bibliography{ref}

\newpage
\appendix
\section{Pseudo-code for NCL}

\SetKw{KwIIn}{\textbf{in}}
\begin{algorithm}
    \caption{Neighborhood-enriched Constrastive Learning~(NCL)}
    \label{alg:ncl}
    \KwIn{bipartite graph $\mathcal{G} = \{ \mathcal{U} \cup \mathcal{I}, \mathcal{E} \}$, training dataset $\mathcal{X}$, number of clusters $K=\{k_m\}_{m=1}^{2M}$, learning rate $\alpha$;}
    \KwOut{user and item representations $\{\bm{z}_u, \bm{z}_i\}$\;}
    Initialize: random initialize user embeddings $\bm{e}_u$ and item embeddings $\bm{e}_i$ \;

    \While{Not Convergence}
    {
    \tcp{E-step} 
    
     \For{m=1 \emph{\KwTo} M}
        {
       $ \bm{c}^u_k = \text{k-means}(\bm{e}_u, k_m)$ ;  \tcp*[h]{k$^{th}$ user prototype} \\
       $ \bm{c}^i_k = \text{k-means}(\bm{e}_i, k_{m+M})$ ; \tcp*[h]{k$^{th}$ item prototype} \\
       }
      \tcp{M-step}
      \For(\tcp*[h]{load minibatch data}){x \KwIIn  \text{Dataloader}($\mathcal{X}$)}
      {
        $\bm{z}_u, \bm{z}_i = \text{GraphConv}(\mathcal{G}, \bm{e}_u, \bm{e}_i) $; \tcp*[h]{forward propagation} \\
        Calculate Loss $\mathcal L(\bm{z}_u, \bm{z}_i, \bm{c}^u, \bm{c}^i)$; \\
        $\bm{z}_u = \bm{z}_u - \alpha \frac{\partial \mathcal L}{\partial {\bm{z}_u}}$; \\
        $\bm{z}_i = \bm{z}_i - \alpha \frac{\partial \mathcal L}{\partial {\bm{z}_i}}$; \tcp*[f]{back propagation} \\
      }
    }
    $\bm{z}_u, \bm{z}_i = \text{GraphConv}(\mathcal{G}, \bm{e}_u, \bm{e}_i) $; \\
    
    return $\bm{z}_u, \bm{z}_i$
\end{algorithm}

\section{Case Study on Selected Neighbors}
\begin{figure}[!h]
    \centering
    \includegraphics[width=0.44\textwidth]{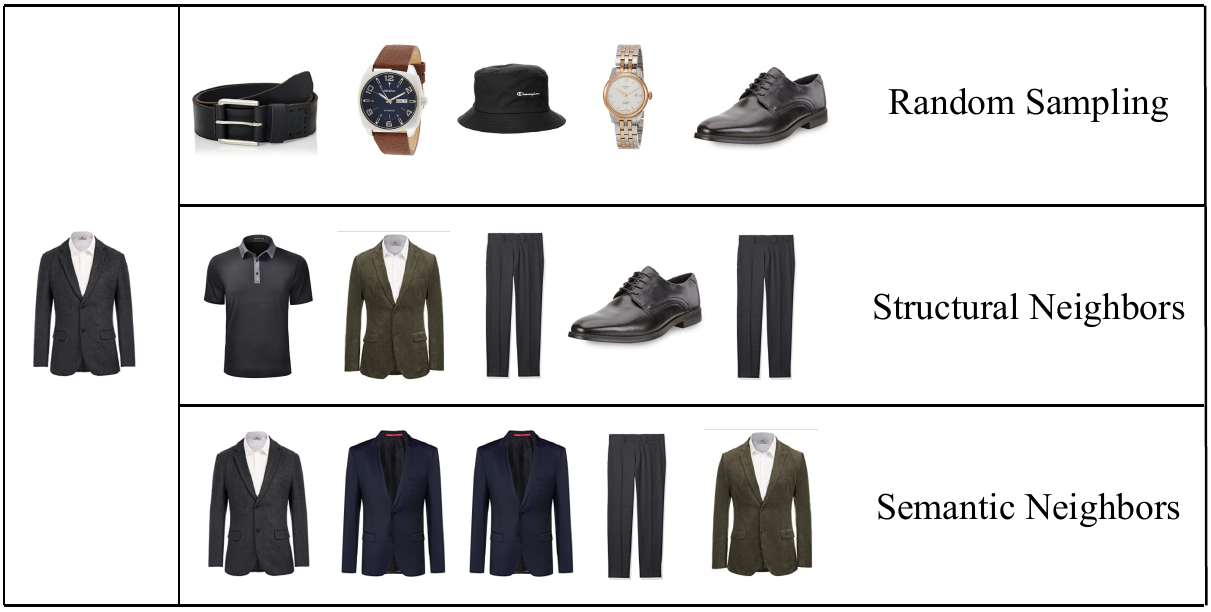}
    \vspace{-0.1in}
    \caption{Case study of the contrastive items sampled from random and proposed structural and semantic neighbors.}
    \label{fig:case}
\end{figure}

To further analyze the difference between structural neighbors and semantic neighbors, we randomly select a central item on Alibaba-iFashion dataset and extract its structural neighbors and semantic neighbors, respectively.
For the two types of neighbors extracted, we count the number of items in each category, respectively.
The number is normalized and visualized in Fig.~\ref{fig:case}. For comparison, we also report the collection of randomly sampled items.
As shown in the figure, the randomly sampled neighbors are uncontrollable, which astrict the potential of contrastive learning.
Meanwhile, the proposed structural and semantic neighbors are more related, which are more suitable to be contrastive pairs.

\end{document}